\documentclass[review, jog]{igs}

\usepackage[utf8]{inputenc}
\usepackage{mathabx}
\usepackage[cal=dutchcal]{mathalpha}
\usepackage{graphicx}
\usepackage{siunitx}
\usepackage{xcolor}
\usepackage{xfrac}
\usepackage{upgreek}

\begin{document}

\title[Time-domain Diffuse Optics for Snow]{Measurement of Snowpack Density, Grain Size, and Black Carbon Concentration Using Time-domain Diffuse Optics}




\author[Henley and others]{Connor HENLEY,$^{1,2}$
  Joseph L. HOLLMANN,$^2$ Colin R. MEYER,$^3$ 
  Ramesh RASKAR$^1$}

\affiliation{%
$^1$MIT Media Lab, Massachusetts Institute of Technology,
  75 Amherst Street, Cambridge, MA, USA\\
$^2$The Charles Stark Draper Laboratory, Inc.,
  555 Technology Square, Cambridge, MA, USA\\
$^3$Thayer School of Engineering, Dartmouth College,
  15 Thayer Drive, Hanover, NH, USA\\
  Correspondence: Connor Henley
  \email{co24401@mit.edu}}

\begin{frontmatter}

\maketitle

\begin{abstract}
Diffuse optical spectroscopy (DOS) techniques aim to characterize scattering media by examining their optical response to laser illumination. Time-domain DOS methods involve illuminating the medium with a laser pulse and using a fast photodetector to measure the time-dependent intensity of light that exits the medium after multiple scattering events. While DOS research traditionally focused on characterizing biological tissues, we demonstrate that time-domain diffuse optical measurements can also be used to characterize snow. We introduce a model that predicts the time-dependent reflectance of a dry snowpack as a function of its density, grain size, and black carbon content, and we develop an algorithm that retrieves these properties from measurements at two wavelengths. To validate our approach, we use a two-wavelength lidar system and measure the time-dependent reflectance of snow samples with varying properties. Rather than measuring direct surface returns, our system captures photons that enter and exit the snow at different points, separated by a small distance (4-10cm).   We find strong, linear correlations between our retrievals of density and black carbon concentration, and ground truth measurements.  Although the correlation is not as strong, we also find that our method is capable of distinguishing between small and large grains.
\end{abstract}


\end{frontmatter}

\section{Introduction}



Snow is composed of transparent ice grains that absorb light very weakly at visible wavelengths \citep{warren2019}.  Because of this, photons that enter a snowpack will typically scatter many times off of a large number of ice grains before they either exit the medium or get absorbed.  The study of snow optics has historically focused on the interaction of snow with \emph{sunlight}, as understanding this interaction is essential to understanding snow cover's contribution to the Earth's climate \citep{henderson2018} and for forecasting snow melt \citep{painter2010}, among other things.  A key goal of snow optics has been to predict spectral albedo as a function of snowpack properties---such as grain size, which determines the probability that a photon will be absorbed at each scattering event \citep{ww1}, and the concentration of light absorbing particles such as dust, black carbon, or algae that are mixed into the snow \citep{WW2, skiles2018}.

The development of accurate spectral albedo models has, in turn, led to the development of optical sensing methods that retrieve grain size \citep{nolin2000, gallet2009} and LAP concentrations \citep{zege2011, painter2012} from spectral albedo measurements.  These methods, while useful, have limitations.  Snowpack albedo is largely independent of important properties such as snow density \citep{ww1}.  Furthermore, spectral albedo measurements usually require passive illumination by sunlight, and as such cannot be used to retrieve snow properties at night and for several months of the year in polar regions.  Albedo models developed for solar illumination assume steady-state illumination that is collimated, diffuse, or a mixture of the two.  As such, they cannot fully model lidar waveform measurements, which consist of the time-dependent optical response of snowpack to focused, pulsed illumination.


Over the past few decades, in parallel with advances in snow optics, the biomedical optics community has developed a suite of techniques for characterizing biological tissue, which, like snow, is also a highly scattering medium.  Collectively, these methods are referred to under the umbrella term of \emph{diffuse optical spectroscopy} (DOS) \citep{durduran2010}, which refers to the fact that the propagation of photons within the scattering medium is modeled using the \emph{diffusion approximation} to the radiative transfer equation \citep{welch95}, and to the fact that multi-wavelength illumination is frequently used (although this is not required).  In DOS techniques the tissue is probed with a focused laser source that can be time-modulated, frequency-modulated, or continuous-wave.  Measurements of the tissue's optical response are then used to estimate its optical properties, such as the tissue's absorption coefficient or effective scattering coefficient.  These optical properties, in turn, can be related to clinically useful properties of the tissue such as blood oxygenation \citep{sevick1991}, organelle size \citep{li2008}, and the concentrations of water, lipids, and collagen \citep{quarto2014}.  DOS has also been applied in non-clinical settings for the inspection of produce \citep{nicolai2014}, and for characterizing porous materials such as wood \citep{bargigia2013} and pharmaceutical tablets \citep{johansson2002}. 


Because snow is also a highly scattering medium, many of the results from diffuse optical spectroscopy can be adapted to the characterization of snowpack properties.  Despite this, the adoption of diffuse optics concepts in the snow sensing community has been limited.  \cite{varnai2007} proposed that the spatial spread of diffused laser light could be used to determine snow and sea ice thickness.   \cite{smith2018} noted that the multiple scattering of green laser light within a snowpack should result in biases in lidar altimetry measurements.  They used a combination of diffusion theory and Monte-carlo modeling to assess the dependence of this multiple scattering bias on grain size, black carbon concentration, and the choice of surface height retrieval algorithm.  \cite{fair2022} experimentally confirmed predictions of \cite{smith2018} by comparing remotely sensed grain size measurements to biases in snow surface heights retrieved using green (532 nm) lidar beams on IceSat-2 and the Airborne Topographic Mapper (ATM).  As far as we are aware, prior to this work, the only direct application of DOS techniques to retrieve bulk snowpack properties was made by \cite{allgaier2022snow}.  In their work, the snow was illuminated with continuous-wave laser sources at two different wavelengths, and a smartphone camera was used to take images of the spatially resolved, steady-state intensity of light that exited the snowpack after diffusing within the snow.  From these smartphone images, along with an independent \emph{in situ} measurement of the snow's density, the authors were able to retrieve the absorption and effective scattering coefficients of the snowpack, as well as an estimate of the concentration of black carbon within it.  In \cite{ackermann2006}, and in a separate work by \cite{allgaier2022ice}, time-domain diffuse optical measurements were used to estimate the absorption and scattering coefficients of \emph{glacier ice}, which is optically similar to snow when it is rich in air bubbles. A theoretical analysis of diffuse optical spectroscopy applied to glacier ice is provided in \cite{allgaier2021}.  \cite{maffione1998} and \cite{perron2021} used steady-state, spatially-resolved reflectance measurements to measure inherent optical properties of \emph{sea ice}.


In this work we introduce what is, to our knowledge, the first method for characterizing the bulk properties of \emph{snow} that is based on \emph{time-domain} diffuse optical measurements.  Our instrument is effectively a photon-counting lidar system that consists of two pulsed lasers with different wavelengths (one red, one near-infrared), and a single-photon avalanche diode (SPAD) receiver.  Rather than measuring surface returns, which might be used for altimetry, we measure photons that enter the snowpack at a single point on the surface and exit at a second surface point that is displaced from the point of entry by a small distance (4-10 cm).  Through a series of proof-of-principle experiments, we show that our method is capable of retrieving the density (through the ice volume fraction), grain size, and the concentration of light absorbing particles of a dry snowpack, in a non-invasive way.  As far as we are aware, ours is the first method to estimate snowpack \emph{density} using non-invasive optical reflectance measurements.  Although our system uses the same components as a photon-counting lidar, our measurements are effectively \emph{in situ}, as the instrument is always placed within a meter of the snow surface.  However, by demonstrating that important bulk snowpack properties can be retrieved from time-domain measurements of multiply scattered photons, we hope that our work will motivate the future development of techniques that retrieve snowpack properties from remote lidar measurements.  We also believe that time-domain diffuse optical measurements, in general, represent a new frontier for studying the optical properties of snow.



\section{Methods}

\subsection{Diffusion Model}

The propagation of a laser pulse inside a scattering medium is described by the time-dependent radiative transfer equation \citep{welch95}, which models the flow of \emph{radiance} ($\text{W} \text{ m}^\text{-2} \text{ sr}^\text{-1}$) within a medium as a function of space and time.  The scattering medium is described by a scattering coefficient $\mu_s$ ($\text{m}^\text{-1}$), a scattering phase function, an absorption coefficient $\mu_a$ ($\text{m}^\text{-1}$), and the speed of light within the medium $c_*$ (m $\text{s}^\text{-1}$).

Under the \emph{diffusion approximation} to the radiative transfer equation, photons are modeled as particles that ``diffuse'' through a scattering medium via random walks.  This approximation accurately describes situations for which the distance scales considered are much larger than the mean free path of photons within the medium (= $(\mu_a+\mu_s)^{-1}$), and photons are typically scattered many times before they are absorbed ($\mu_s \gg \mu_a$) \citep{welch95}. The photon diffusion equation can be written as follows:

\begin{equation}
\label{eqn:diff}
    \frac{1}{c_*} \frac{\partial}{\partial t} \phi(\mathbf{r}, t) - D \nabla^2\phi(\mathbf{r}, t) + \mu_a \phi(\mathbf{r}, t) = S(\mathbf{r}, t)  .
\end{equation}

A derivation of the photon diffusion equation can be found in \cite{haskell1994}.  Unlike the time-dependent radiative transfer equation, which models the time-evolution of a five-dimensional radiance field, the photon diffusion equation models the lower dimensional quantity of \emph{photon fluence} $\phi(\mathbf{r}, t)$ ($\text{W}\text{ m}^\text{-2}$), which is the integral of radiance over all directions.  Here $D = 3\left[ \mu_a + (1 - g)\mu_s \right]^{-1}$ is the diffusion constant and $g$ is the anisotropy factor of the scattering phase function, which can take values between $-1$ and $1$ depending on whether the medium is primarily backward scattering ($g < 0$), isotropically scattering ($g=0$), or forward scattering ($g>0$).  The variable $S$ represents an isotropic source term.

\begin{figure}
\centering{\includegraphics[width=\linewidth]{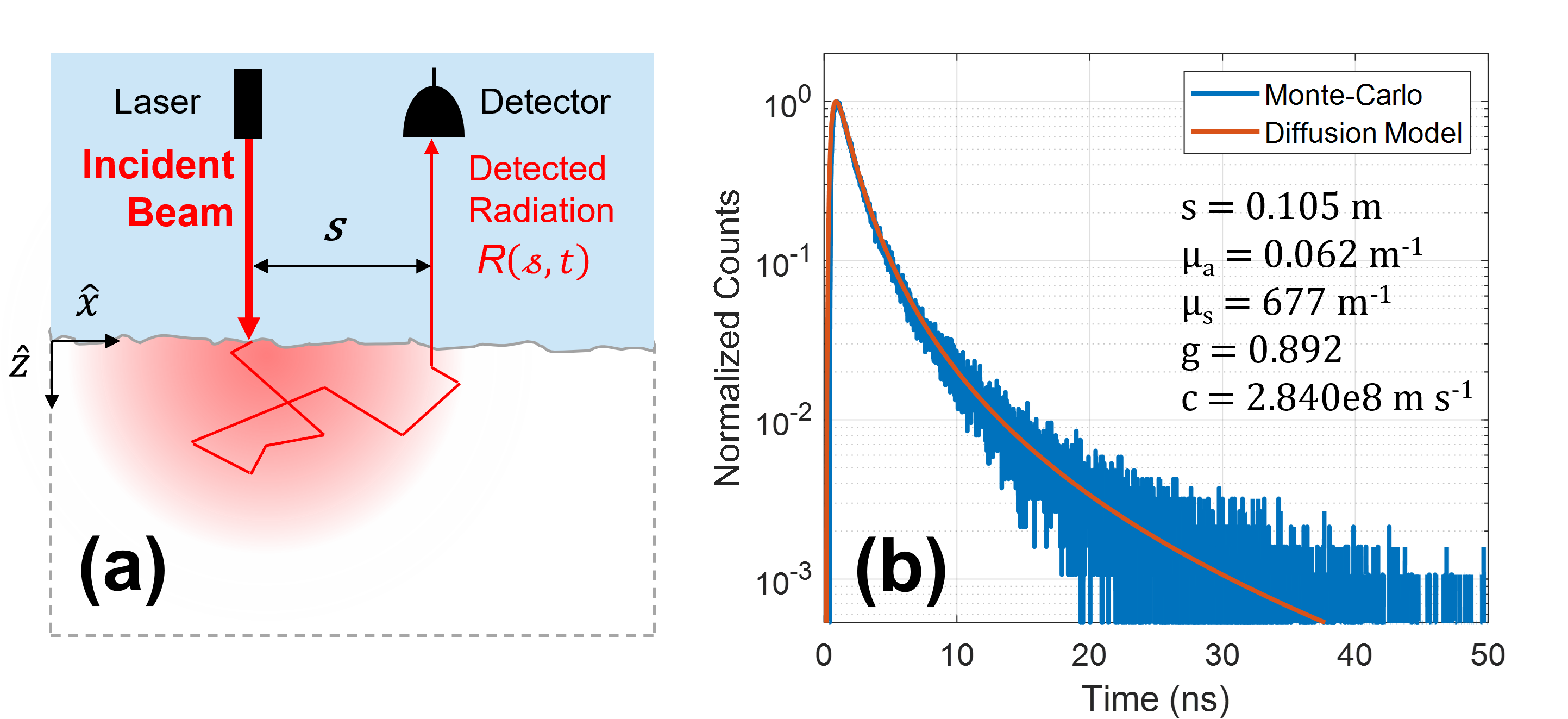}}
\caption{(a) Illustration of the measurement geometry employed in this work.  A point on the snow surface is illuminated by a laser pulse at time $t=0$.  A detector observes the time-dependent intensity of light that exits the snow from a second point at distance $s$ from the laser spot.  (b) Comparison of time-dependent intensity predicted by our model (Eq.~\ref{eqn:R}), to photon time-of-flight measurements generated using a Monte-carlo simulation of a scattering medium with the same properties.  Both curves are normalized to their respective peaks.}
\label{fig:cartoon}
\end{figure}

Crucially, the photon diffusion equation permits analytical solutions when the geometry of the scattering medium is sufficiently simple.  We consider the scenario depicted in Fig.~\ref{fig:cartoon}(a).  Here, the medium is assumed to be semi-infinite and homogeneous.  The medium's surface is illuminated by a pulsed, pencil-beam source at time $t = 0$, and a detector observes the time-dependent intensity of light that exits the medium at a second point that is displaced from the point of illumination by a distance $s$.  \cite{kienle1997} showed that, in this scenario, Eq.~\ref{eqn:diff} can be accurately solved by imposing an \emph{extrapolated boundary condition}, which requires that photon fluence goes to zero along a planar boundary that lies just above the medium's surface.  This yields the following expression for photon fluence inside the medium:


\begin{equation}
\label{eqn:fluence}
    \phi(s, z, t) = \frac{c_*}{(4\pi Dc_*t)^{3/2}}\exp{\left(-\mu_a c t\right)} \left\{ \exp{\left[-\frac{(z-z_0)^2 + s^2}{4 D c_* t}\right]} - \exp{\left[-\frac{(z+z_0+2z_b)^2 + s^2}{4 D c_* t}\right]} \right\} .
\end{equation}


Here $z$ denotes the distance from the surface, going down; $z_0 = [\mu_a + (1-g)\mu_s]^{-1}$ is the transport mean free path of photons within the medium; and $z_b$ denotes the height of the extrapolated boundary.  \cite{haskell1994} proposed a value of $z_b = \frac{1+R_{eff}}{1-R_{eff}}2D$, where $R_{eff}$ is the fraction of photons that are internally reflected at the interface between the scattering medium and the external (non-scattering) medium due to a refractive index mismatch.  Because our ultimate goal is to model the optical response of a snowpack, and because the snow-air boundary of a typical snowpack is not a dielectric interface at optical wavelengths, we assume for this work that $R_{eff}=0$ and hence, $z_b=2D$.

From Eq. \ref{eqn:fluence}, we compute the time dependent radiosity ($\text{W} \text{ m}^\text{-2}$) that exits the surface at position $s$ using Fick's Law \citep{kienle1997}:

\begin{equation}
    \label{eqn:ficks}
    J(s, t) = -D\nabla\phi(s, z, t)\cdot(-\hat{z})|_{z=0} .
\end{equation}

The reflected flux $R$ measured by a detector that observes the medium's surface from a distance can then be described using the following expression:





\begin{equation}
    \label{eqn:R}
    R(s,t) = \frac{\alpha c_*}{3(2\pi)^{3/2}}\frac{z_0^2}{(2Dc_*t)^{5/2}}\exp\left(-\mu_ac_*t -\frac{s^2 + z_0^2}{4Dc_*t}\right)\left[1 + \frac{7}{3}\exp\left(-\frac{10z_0^2}{9Dc_*t}\right) \right] ,
\end{equation}


where $\alpha$ is a constant that encapsulates instrumental parameters such as transmitted laser power, detection efficiency, and the detector's etendue.  We note that we have made liberal use of the substitutions $D = z_0/3$ and $z_b = 2z_0/3$. In deriving Eq.~\ref{eqn:R}, we also assumed that the surface could be accurately described as a Lambertian emitter, which means that the radiance emitted by the surface is independent of the emission angle.  Previous work has relaxed this assumption \citep{kienle1997}.  We found that doing so produced nearly identical results when describing a nadir-pointing detector, but added significant complexity to the model.  For this reason, we elected to use Eq.~\ref{eqn:R}. 

In Fig.~\ref{fig:cartoon}(b) we compare the time-dependent intensity predicted by Eq.~\ref{eqn:R} to simulated photon time-of-flight measurements generated using a Monte-carlo simulation \citep{SnowLiDARMonteCarlo}.  The modeled results match the simulation very closely.  In general, models derived from the diffusion approximation to the radiative transfer equation accurately describe the measurements of photons that arrive at later times ($c_*t\gg z_0$) and larger distances from the laser spot ($s \gg z_0$), as these photons have scattered many times before exiting the medium.

\subsection{Snow Scattering Model}

Our measurement model, defined in Eq.~\ref{eqn:R}, is expressed in terms of three phenomenological parameters---the absorption coefficient $\mu_a$, the effective scattering coefficient $\mu_s'=(1-g)\mu_s$, and the effective speed of light in the medium $c_*$.  We use a scattering model derived from the geometric-optics scattering model of \cite{kokh2004} to define $\mu_a$, $\mu_s'$, and $c_*$ in terms of three physically meaningful snowpack parameters---$v_*$, the fraction of the snowpack volume that is occupied by ice; $r_*$ (m), the grain radius; and $C_{bc}$ (kg $\text{kg}^\text{-1}$), the mass mixing ratio of black carbon in the snowpack.  We note that, for a dry snowpack, the ice volume fraction $v_*$ is readily converted to bulk snowpack density $\rho_*$ (kg $\text{m}^\text{-3}$) via the expression $\rho_* = v_*\rho_{ice} + (1-v_*)\rho_{air}\approx v_*\rho_{ice}$, where $\rho_{ice}$ and $\rho_{air}$ are the intrinsic densities of ice and air, respectively.  We do not consider wet snow in this work.

\subsubsection{Clean Snowpack}

For a dry snowpack that contains optically insignificant concentrations of light absorbing particles, the scattering and absorption coefficients can be written entirely as functions of $v_*$ and $r_*$.  The absorption and effective scattering coefficients are computed as follows: 

\begin{equation}
\label{eqn:mua}
    \mu_a = B\Gamma v_*
\end{equation}

\begin{equation}
\label{eqn:museff}
    \mu'_s = \frac{3}{2}(1-g)\frac{v_*}{r_*} ,
\end{equation}

Here, as in \cite{kokh2004}, the grain radius $r_*$ can be interpreted as the radius of the spherical ice grain that would have the same surface-area-to-volume ratio as the ice-air matrix that comprises the true snowpack.  Explicitly $r_*=3\frac{\langle V\rangle}{\langle \Sigma\rangle}$, where $\langle V\rangle$ is the mean ice grain volume and $\langle \Sigma\rangle$ is the mean ice grain surface area.  The absorption coefficient of pure ice $\Gamma = \frac{4\pi\kappa_{ice}}{\lambda}$, where $\lambda$ corresponds to the light's wavelength, and $\kappa_{ice}$ is the imaginary component of ice's refractive index at that wavelength \citep{warren2008}.  The absorption enhancement factor $B$ accounts for the lengthening of photon path length within the ice phase due to internal reflections.

Although the absorption enhancement parameter $B$ and scattering asymmetry $g$ theoretically depend on grain shape \citep{libois2013}, a recent study by \cite{robledano2023} suggests that these parameters cluster around $B=1.7$ and $g=0.825$ for most real snow samples, so we use those values here.  These values are approximately valid for visible and near-infrared wavelengths (400nm to 14000nm) \citep{robledano2023}.


\begin{figure}
\centering{\includegraphics[width=\linewidth]{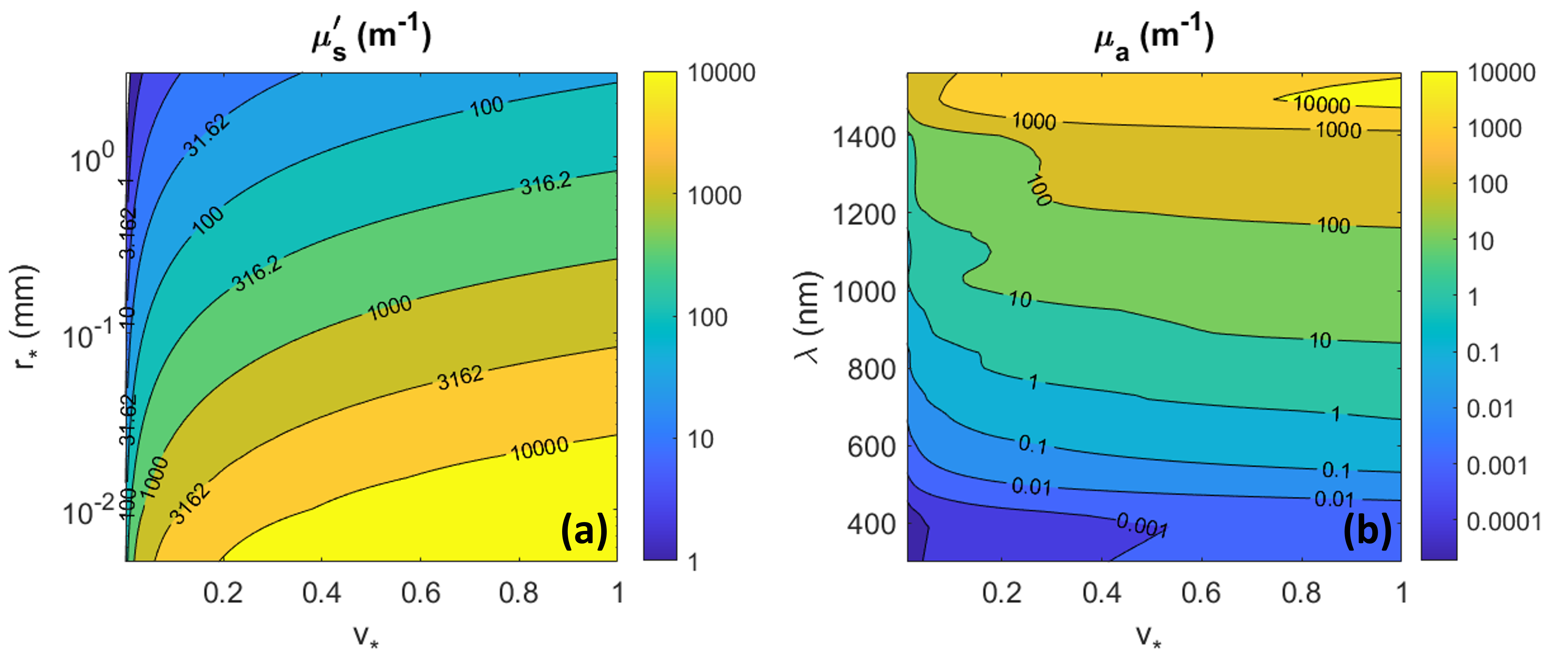}}
\caption{(a) Effective scattering coefficient $\mu_s'$ ($\text{m}^\text{-1}$) as a function of ice volume fraction $v_*$ (unitless) and grain radius $r_*$ (mm). (b) Absorption coefficient $\mu_a$ ($\text{m}^\text{-1}$) of clean snow as a function of ice volume fraction $v_*$ (unitless) and wavelength $\lambda$ (nm).} 
\label{fig:muSmuA}
\end{figure}


In Fig.~\ref{fig:muSmuA}(a) we visualize the range of values for $\mu_s'$ obtained across a domain of feasible grain sizes and ice volume fractions. Figure \ref{fig:muSmuA}(b) shows the dependence of $\mu_a$ on ice volume fraction and wavelength.  Unlike $\mu_s'$, the absorption coefficient depends strongly on wavelength, and varies by more than an order of magnitude between the red ($\lambda = $ 640 nm) and near infrared ($\lambda = $ 905 nm) wavelengths used in this study.

\subsubsection{Effective Speed of Light in Snow}

The last parameter to calculate is the effective speed of light within the snowpack $c_*$.  In many problems that involve light propagation in a scattering medium, light's speed is treated as a constant that can be computed beforehand if the medium's index of refraction is known.  This approach does not work for snow, which is a heterogeneous mixture of two materials---ice and air---that have markedly different refractive indices and that may be mixed at any ratio.  

\begin{figure}
\centering{\includegraphics[width=0.5\linewidth]{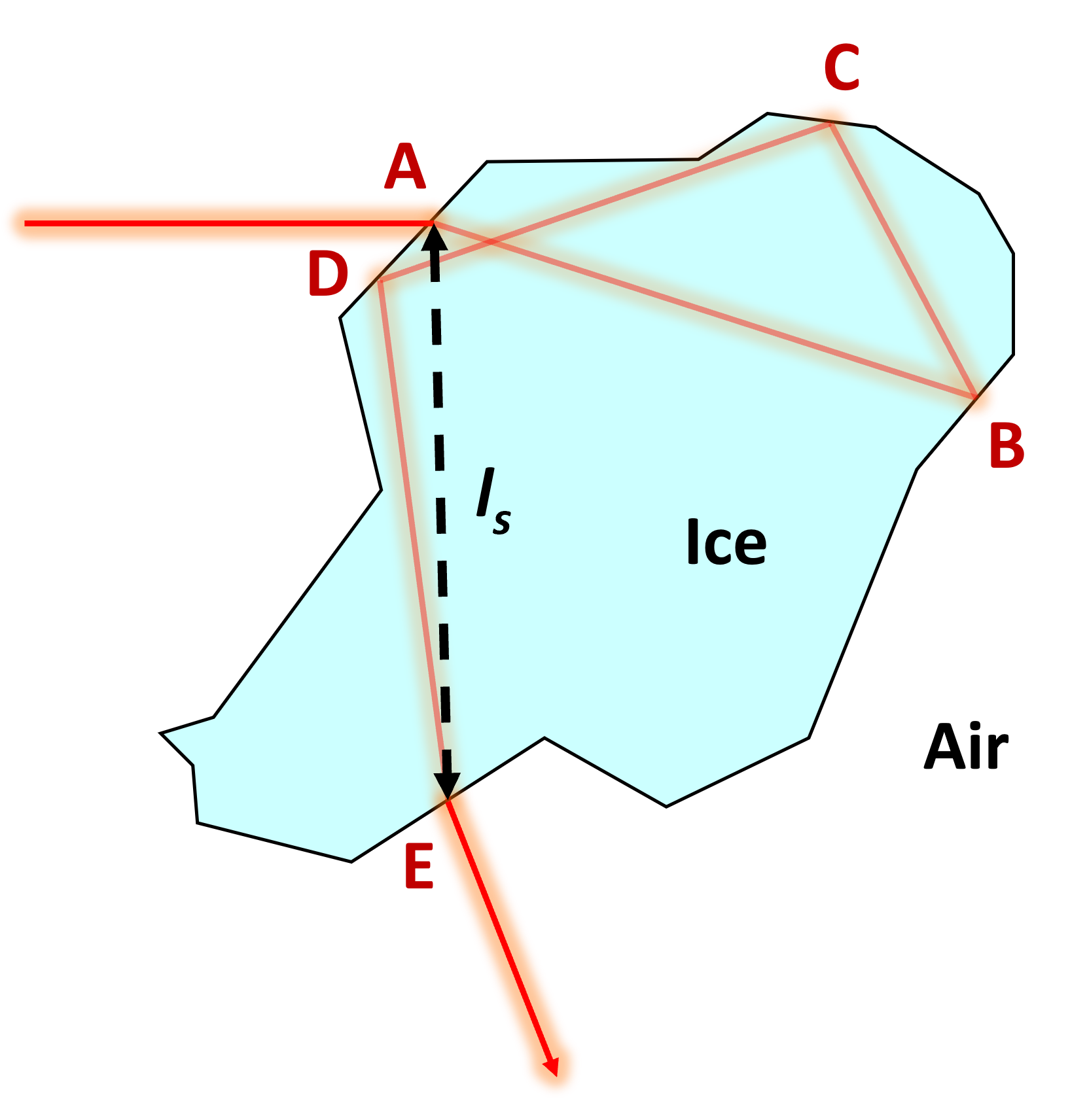}}
\caption{A comparison of the true path (red) traveled by a photon through an ice grain, including internal reflections, to the effective transportation path (black, dashed) of length $l_s$ that is implicitly assumed by our absorption and effective scattering coefficient models.}
\label{fig:c_cartoon}
\end{figure}

The geometric scattering model proposed by \cite{kokh2004}, which is the source of Eqs. ~\ref{eqn:mua} and \ref{eqn:museff}, implicitly defines the distance that a photon travels through an ice grain as the distance (i.e. $l_{ice}$) of the chord that connects the points at which a photon enters (point A in Fig.~\ref{fig:c_cartoon}) and exits (point E) the grain.  This effective ``transportation distance'' differs from the \emph{true} distance (i.e. $l_{ice}'$) traveled through the grain if the photon is internally reflected (e.g.~at points B, C and D) before it exits the grain.  The absorption enhancement factor $B$ is approximately equal to the expected ratio between the true distance and effective transportation distance, i.e. $l_{ice}' \approx Bl_{ice}$ \citep{libois2019}.

An effective light speed model that is consistent with Eqs.~\ref{eqn:mua} and \ref{eqn:museff} must describe the average speed at which light advances along this effective transportation path, which is equivalent to the true photon path in the air phase, but shorter than the true photon path in the ice phase by a factor of $B$.  If a photon travels along an effective transportation path of length $L$, on average that path will pass through $(1-v_*)L$ of air and $v_*L$ of ice.  The time $T$ require to traverse this path is

\begin{equation}
    T = \frac{(1-v_*)L}{c_0} + \frac{n_{ice} B v_* L}{c_0} ,
\end{equation}

where $n_{ice}$ is the real component of the refractive index of ice \citep{warren2008} and $c_0$ is the speed of light in air (where it's assumed that $n=1$).  The travel time within the ice phase has been increased by a factor $B$ to account for the difference between the true and effective transportation path lengths.  Dividing $L$ by $T$ leaves us with

\begin{equation}
    \label{eqn:c}
    c_* = \frac{c_0}{1 + (n_{ice}B-1)v_*} .
\end{equation}

We stress that the effective light speed defined here is \emph{lower} than the mean speed of light computed with respect to the lengths of true photon paths through snow, which due to internal reflections can include jagged paths within grains.  An expression for this true mean light speed was computed by \cite{libois2019}, and is equal to the effective light speed of Eq.~\ref{eqn:c}, multiplied by a factor of $[1 + (B-1)v_*]$.

\subsubsection{Effect of Light Absorbing Impurities}

Ice is an exceptionally weak absorber of light at visible wavelengths \citep{warren2019}.  As such, the absorption of visible light within a snowpack can be enhanced significantly---even dominated---by the presence of trace concentrations of more absorptive substances.  This has the important effect of reducing snowpack albedo, which increases radiative forcing on the snow surface and subsequently enhances snow melt and metamorphism and can also influence the local climate \citep{skiles2018}.  For our purposes, the presence of small concentrations of LAPs can increase the absorption coefficient of a snowpack considerably---thus rendering Eq.~\ref{eqn:mua}, our model for clean snowpack absorption, insufficient.  Globally, radiative forcing from LAPs is dominated by black carbon, mineral dust, organic or ``brown'' carbon, and snow algae \citep{skiles2018}.  Here we assume that absorption by LAPs is dominated by black carbon, but note that our model could be extended to include other types of particles by modifying the LAP absorption spectrum used here.

According to \cite{flanner2012}, between 32-73\% of the black carbon in global surface snow is embedded within ice grains (or ``internally mixed''), with the remainder being external to those grains (``externally mixed'') in the air phase.  The elongated paths followed by photons within ice increases the probability that photons will interact with internally mixed black carbon particles.  As a consequence, internally mixed black carbon has an outsized impact on snow absorption and albedo, relative to externally mixed black carbon \citep{flanner2012}.  Models for snow's absorption coefficient that consider the mixing state of black carbon have been proposed \citep{liou2014, dombrovsky2020}, however these models typically require idealized grain shapes such as spheres---which do not accurately represent real snow---and assign highly non-linear dependencies on black carbon concentration that are grounded in electromagnetic theory \citep{dombrovsky2020} or stochastic simulations \citep{liou2014}.

Here we propose a simple geometric optics model for the additional absorption due to black carbon that can be computed from the bulk density of black carbon particles embedded inside ice grains $\rho_{bc}^{in}$ (kg $\text{m}^\text{-3}$), the bulk density of black carbon particles external to the grains $\rho_{bc}^{out}$ (kg $\text{m}^\text{-3}$), and the wavelength-dependent mass absorption efficiency $\textrm{MAE}_{bc}$ ($\text{m}^\text{2}\text{ kg}^{-1}$) of the black carbon particles \citep{grenfell2011}.  Under this model, the presence of black carbon in a snowpack alters its properties primarily by adding an extra term to the absorption coefficient, i.e. $\mu_a^{snow} = \mu_a^{ice} + \mu_a^{bc}$.  Our proposed model is written as follows:



\begin{equation}
    \label{eqn:muabc}
    \mu_a^{bc} = \textrm{MAE}_{bc}\left[(1-v_*)\rho_{bc}^{in}+Bv_*\rho_{bc}^{out}\right] = \textrm{MAE}_{bc}\rho_{ice}v_*\left[ (1-v_*)C_{bc}^{out} + Bv_*C_{bc}^{in}\right] .
\end{equation}

Here $B$ is the absorption enhancement factor of the snowpack.  On the right side of Eq~\ref{eqn:muabc}, we replace $\rho_{bc}^{in}$ and $\rho_{bc}^{out}$ with the products of the intrinsic density of ice ($\rho_{ice} = $ 916.5 kg $\text{m}^\text{-3}$), the snowpack's ice volume fraction $v_*$, and the mass mixing ratios $C_{bc}^{in}$ and $C_{bc}^{out}$ of internally and externally mixed black carbon, respectively. 

To simplify our model further, we assume that the black carbon is evenly mixed, i.e. $C_{bc}^{in} = C_{bc}^{out}$.  We then combine Eqs.~\ref{eqn:mua} and \ref{eqn:muabc} to obtain a complete expression for the snowpack absorption coefficient:

\begin{equation}
    \label{eqn:muatot}
    \mu_a = B\Gamma v_* + \textrm{MAE}_{bc}\rho_{ice}C_{bc}v_*\left[ 1 + (B-1)v_*\right] .
\end{equation}

Following the example of \cite{doherty2014}, we model the wavelength dependence of $\textrm{MAE}_{bc}$ using a power law spectrum:

\begin{equation}
    \label{eqn:angst}
    \textrm{MAE}_{bc}(\lambda) = \textrm{MAE}_{bc}(\lambda_{ref})\left(\lambda_{ref}/\lambda\right)^\textrm{{\AA}}, 
\end{equation}

that has an {\AA}ngstrom coefficient {\AA} $=1.1$ and is referenced to $\textrm{MAE}_{bc}(\lambda_{ref}) = 6500$ $\text{m}^\text{2}\text{ kg}^{-1}$, where $\lambda_{ref}=600$ nm.

\begin{figure}
\centering{\includegraphics[width=0.8\linewidth]{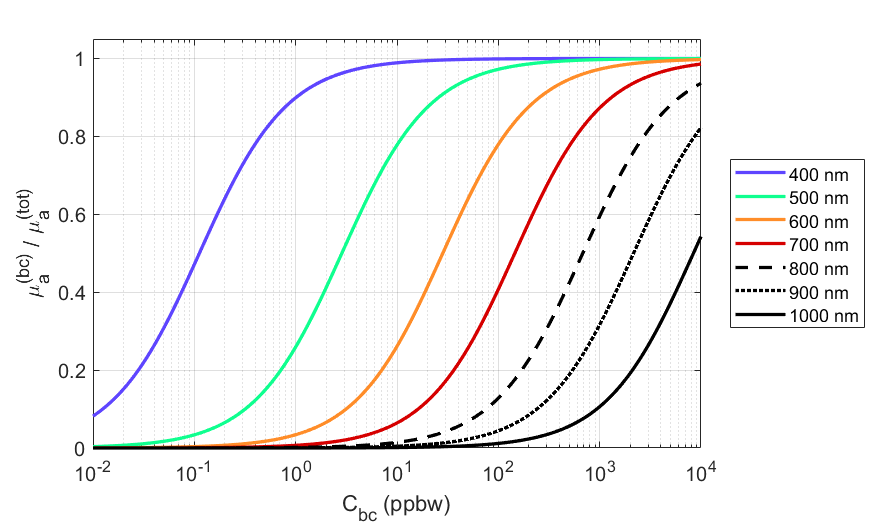}}
\caption{The ratio of light absorption due to black carbon (Eq.~\ref{eqn:muabc}) to total absorption by the snowpack (Eq.~\ref{eqn:muatot}) for a range of wavelengths.  Ratio computation assumes ice volume fraction $v_* = 0.3$.}
\label{fig:muabc_v_tot}
\end{figure}

Fig.~\ref{fig:muabc_v_tot} illustrates that, for a fixed $C_{bc}$, the fraction of absorption attributable to black carbon in snow depends strongly on the wavelength of light that interacts with the snowpack.  We plot the ratio of absorption due to black carbon (using Eq.~\ref{eqn:muabc}) to the total absorption (from Eq.~\ref{eqn:muatot}) for a selection of wavelengths that range from 400 nm (blue) to 1000 nm (near infrared).  In computing these ratios, we assume an ice volume fraction of $v_*=0.3$. For blue light, our model suggests that absorption is entirely dominated by just 1 part per billion by weight (ppbw) of black carbon, which is comparable to mass mixing ratios found in Greenlandic snow \citep{warren2019}.  In contrast, at 1000 nm, absorption from black carbon only eclipses ice absorption for mass mixing ratios above 7500 ppbw---a very high level of soot that would cause the snow to appear visibly grey.  This decreased sensitivity at longer wavelengths is not caused by the decreased absorption of black carbon at these wavelengths, but rather by the increased absorption efficiency of ice.

\subsubsection{Effect of Snow Properties on Time-domain Response}

Having obtained expressions that relate $\mu_a$, $\mu_s'$, and $c_*$ to the grain size, ice volume fraction, and black carbon concentration of a dry snowpack, we can now develop an understanding of how changes to $v_*$, $r_*$, and $C_{bc}$ affect the snowpack's time-domain optical response.  Upon inspection of Eq.~\ref{eqn:R}, we see that the shape of a snowpack's transient response is primarily controlled by $\mu_ac_*$, which determines the rate of decay of the signal's tail; $2Dc_*$, which can be interpreted as the rate at which a Gaussian cloud of diffusing photons expands over time, and which controls the position of the signal's peak; and $z_0^2$, which influences the shape of the response at the earliest arrival times, but in practice has little effect when $s\gg z_0$.

\begin{figure}
\centering{\includegraphics[width=\linewidth]{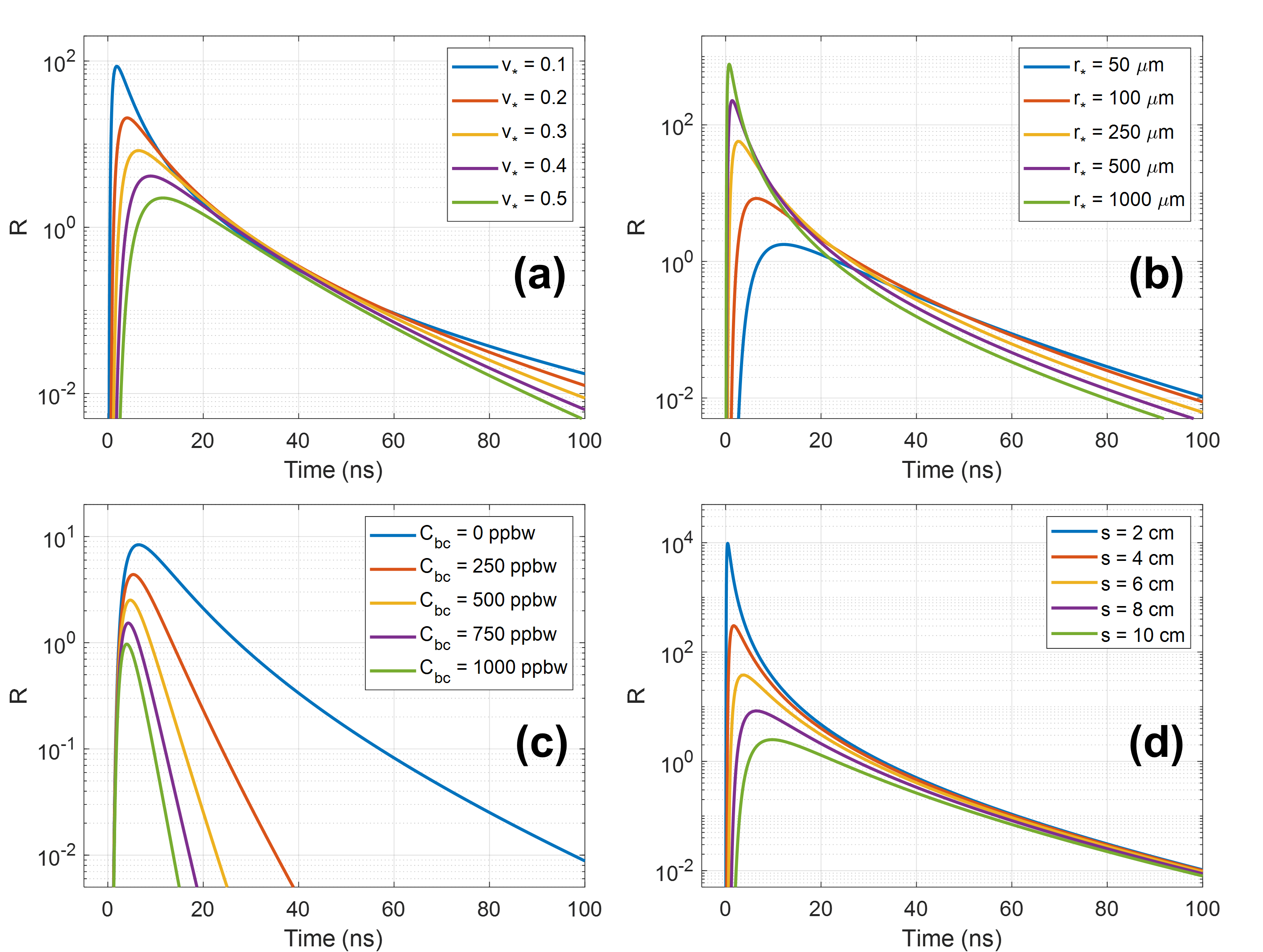}}
\caption{Plots of predicted time-dependent flux measured by a detector observing snow following illumination by a laser pulse ($\lambda = $ 640 nm).  Curves were produced using Eq.~\ref{eqn:R} and $\mu_a$, $\mu_s'$, and $c_*$ were computed from $v_*$, $r_*$, and $C_{bc}$ using Eqs.~\ref{eqn:muatot}, \ref{eqn:museff}, and \ref{eqn:c}, respectively. (a) Ice volume fraction $v_*$ is varied. $r_*$ = 100 $\upmu$m, $C_{bc}$ = 0 ppbw, and $s$ = 8 cm. (b) Grain radius $r_*$ is varied. $v_*$ = 0.3, $C_{bc}$= 0 ppbw, and $s$ = 8 cm. (c) Impurity concentration $C_{bc}$ is varied. $v_*$ = 0.3, $r_*$ = 100 $\upmu$m, and $s$ = 8 cm. (d) Detector focus position $s$ is varied. $v_*$ = 0.3, $r_*$ = 100 $\upmu$m, and $C_{bc}$ = 0 ppbw.}
\label{fig:var}
\end{figure}

The exponential decay rate, $\mu_ac_*$, depends on $v_*$ and $C_{bc}$.  On the other hand, $2Dc_*$ and $z_0^2$ primarily depend on the medium’s scattering coefficient, which in turn depends on the ratio $v_*/r_*$.  These effects are visualized in Figs \ref{fig:var}(a), (b), and (c), where we plot the predicted transient response curves for snowpack with varying $v_*$, $r_*$, and $C_{bc}$.  For these curves, the snow is probed with red (640 nm) light, and the position of the detector's focus spot is fixed at $s = $ 8 cm.  

In Fig.~$\ref{fig:var}(a)$ we see that for a clean snowpack ($C_{bc}=0$), as $v_*$ is increased while $r_*$ is held constant, the slope of the signals' exponential tail becomes more steep as light is absorbed by the medium more quickly.  The arrival time of the signal peak is also pushed back because tighter packing of the ice grains reduces the distance between photon scattering events, thus reducing the rate at which light diffuses within the medium.  

In Fig.~\ref{fig:var}(b), $r_*$ is varied while $v_*$ is held constant and again $C_{bc}=0$.  As grain size increases, grains must be spaced further apart to maintain the same density, thus increasing the rate of diffusion within the medium.  As such, for a fixed snow density and source-detector separation, the peak of the diffusion signal will arrive earlier, and will be more intense, when the grains are large.  

In Fig.~\ref{fig:var}(c), $v_*$ and $r_*$ are held fixed while $C_{bc}$ is varied. Black carbon content only influences the absorption coefficient of the snowpack, and so increasing $C_{bc}$ steepens the exponential decay rate $\mu_ac_*$.  At the probing wavelength of 640 nm this effect is quite dramatic when compared to the comparable influence of ice volume fraction on the exponential decay rate, shown in Fig.~\ref{fig:var}(a).

In Fig.~\ref{fig:var}(d), $v_*$, $r_*$, and $C_{bc}$ are held fixed and the detector focus position $s$ is varied.  As $s$ increases, the signal peak arrives later and becomes more faint.  However, as time passes, all signals converge as light spreads within the medium and the distribution of emitted photons becomes nearly uniform across the observed region.

\subsection{Algorithm}

\begin{figure}
\centering{\includegraphics[width=0.9\linewidth]{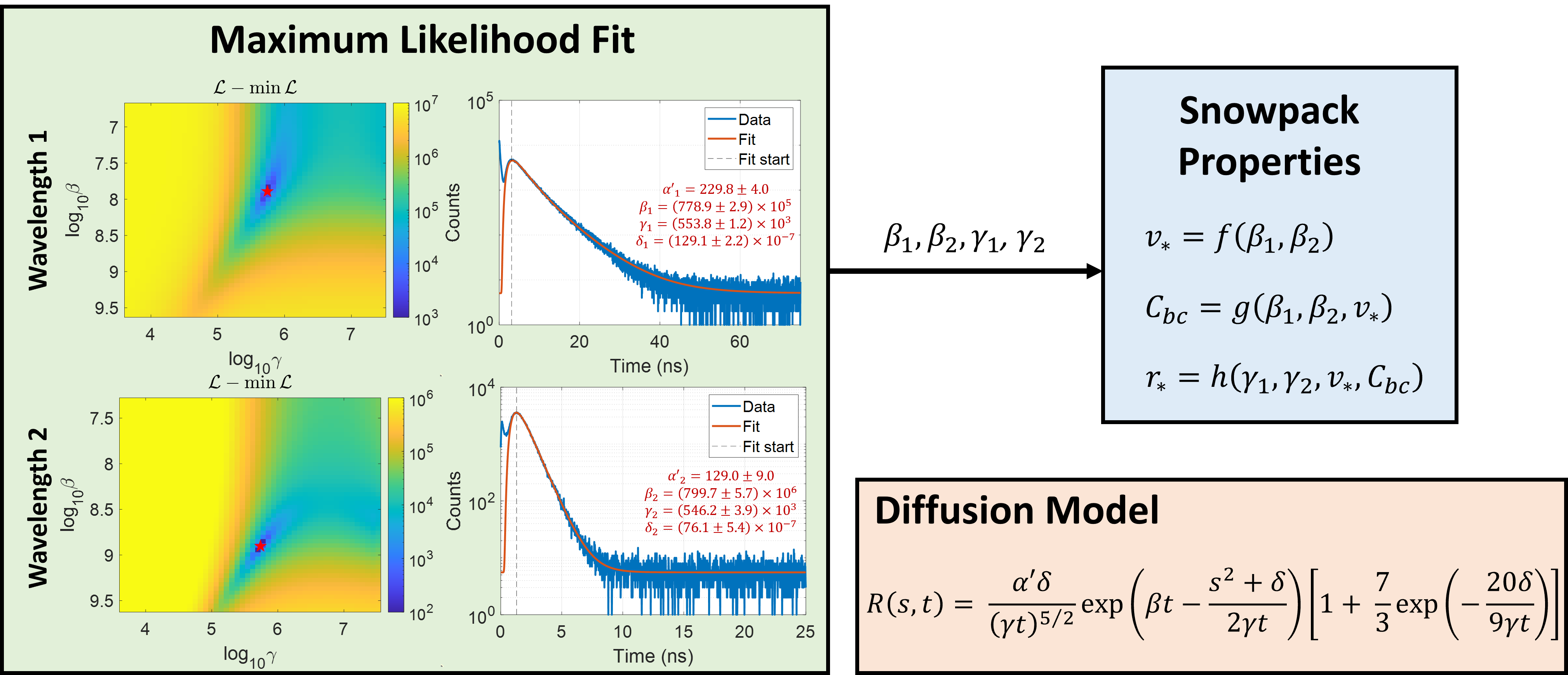}}
\caption{(Left) A diffusion model (lower right) is fit to photon time-of-flight histograms measured at two wavelengths.  Fit parameters $\alpha'$, $\beta$, $\gamma$, and $\delta$ are determined using a grid search algorithm that minimizes a negative Poisson log-likelihood function.  (Top right) Snowpack properties $v_*$, $r_*$, and $C_{bc}$ are computed directly from parameters $\beta_1$, $\beta_2$, $\gamma_1$, and $\gamma_2$ by evaluating analytical expressions (Eqs.~\ref{eqn:v}, \ref{eqn:Cbc}, and \ref{eqn:r})}
\label{fig:algo}
\end{figure}

We fit functions of the same form as Eq.~\ref{eqn:R} to two photon time-of-flight histograms---each measured using a different laser wavelength.  We use a grid search to find the fit parameters that minimize a negative Poisson log-likelihood function that properly accounts for photon count statistics.  The parameters of the fitted curves are then used to compute the snowpack properties $v_*$, $r_*$, and $C_{bc}$.  A visualization of our retrieval algorithm is shown in Fig.~\ref{fig:algo}.

\subsubsection{Fit Parameterization}

We re-parameterize Eq.~\ref{eqn:R} in terms of the fitting parameters $\alpha' = \dfrac{\alpha c_*}{3(2\pi)^{3/2}}$, $\beta = \mu_a c_*$, $\gamma = 2Dc_*$, and $\delta = z_0^2$.  This allows for the model to be expressed in simplified form:

\begin{equation}
    \label{eqn:Rp}
    R(s,t) = \alpha'\frac{\delta}{(\gamma t)^{5/2}}\exp\left(-\beta t -\frac{s^2 + \delta}{2\gamma t}\right)\left[1 + \frac{7}{3}\exp\left(-\frac{20\delta}{9\gamma t}\right) \right] .
\end{equation}

Although Eq.~\ref{eqn:Rp} appears to have four degrees of freedom, only the exponential decay rate parameter $\beta$ and the spatial spread rate $\gamma$ are used to estimate snowpack properties in practice.  Interpreting the scaling constant $\alpha'$ requires precise calibration of the instrument and measurement geometry which is challenging in practice and which we did not attempt.  Eq.~\ref{eqn:Rp} also depends only weakly on the squared source depth $\delta$, to the point that accurate estimates of $\delta$ are almost never obtained.

The spatial spread rate $\gamma$ depends primarily on $\mu_s'$ and $c_*$, which are largely independent of the measurement wavelength for visible and near-infrared light.  Altogether, this means that $n+1$ independent parameters can be retrieved from measurements taken at $n$ wavelengths.  If absorption due to light absorbing particles is known to be insignificant (i.e. $\mu_a^{bc} \ll \mu_a^{ice}$), then $v_*$ and $r_*$ can be retrieved from measurements at a single wavelength.  Otherwise, retrieving $v_*$, $r_*$, and $C_{bc}$ requires measurements at two or more wavelengths.

\subsubsection{Maximum Likelihood Estimate of Model Parameters}

The number of counts in a histogram timing bin centered at $t_i$ is assumed to be a Poisson random variable with a rate parameter $x_i$, defined as

\begin{equation}
    \label{eqn:poissonrate}
    x_i = R(s, t_i ; \mathbf{\Theta}) + \eta = \alpha'r_i + \eta , 
\end{equation}

where $R$ is the flux predicted by Eq.~\ref{eqn:Rp} at position $s$, time $t_i$, and for parameters $\mathbf{\Theta} = \{ \alpha', \beta, \gamma, \delta \}$.  The rate of background counts produced by ambient light, detector dark counts, and detector afterpulsing \citep{zappa2007} is denoted by $\eta$, and is assumed to be constant with respect to time.  The variable $r_i$ denotes the normalized predicted flux, for which $\alpha'=1$.

The probability of observing a vector of time-binned photon counts $\mathbf{y}$ given a vector of predicted count rates $\mathbf{x}$ is 

\begin{equation}
    \label{eqn:poisson}
    P(\mathbf{y}|\mathbf{x}) = \prod^N_\Delta \frac{x_i^{y_i}e^{-x_i}}{y_i!}
\end{equation}

where $N$ denotes the total number of timing bins in the histogram and $\Delta$ is the starting bin for the curve fit.  We seek to find the parameters $\alpha'$, $\beta$, $\gamma$, $\delta$, and $\eta$ that minimize the negative Poisson likelihood

\begin{equation}
    \label{eqn:loglike}
    \mathcal{L}(\mathbf{x}, \mathbf{\Theta}, \eta | \mathbf{y}) = -\ln{P(\mathbf{y}|\mathbf{x})} = \sum^N_\Delta x_i - y_i\ln{x_i} + \ln{y_i!} .
\end{equation}

We find the set of parameters that minimizes the negative log likelihood using a grid search.  To reduce the dimensionality of the search, we first estimate $\eta$ by computing the mean number of counts in a designated set of noise bins that reliably contains effectively zero non-background counts.  For any combination of $\beta$, $\gamma$, and $\delta$, the scaling term $\alpha'$ can then be estimated using the expression $\Bar{\alpha}' = \sum^N_\Delta \left(y_i - \eta \right)/\sum^N_\Delta r_i$.


We can thus define a three-dimensional search area that contains all feasible values of $\beta$, $\gamma$, and $\delta$.  The feasible range for $\delta$ is tightly constrained to $\left( \dfrac{3\gamma}{2c_0} \right)^2 < \delta < \left( \dfrac{3n_{ice}\gamma}{2c_0} \right)^2$, which allows for a coarse fit to be obtained using an effectively two-dimensional search.

We perform a sequence of nested searches---we first obtain a coarse fit, then define a small search range around the fitted parameters and repeat the search using a smaller grid cell size.  This procedure is iterated until a fit with the desired precision is obtained.  Our fitting algorithm was implemented in MATLAB on a Lenovo Thinkpad T590 laptop with 16GB of RAM.  Run time per fit was typically 56 seconds for 640 nm histograms, and 19 seconds for 905 nm histograms (which had fewer timing bins). Curve fits obtained using our algorithm are shown in Fig.~\ref{fig:algo}.  We estimated the uncertainty in the retrieved values of $\beta$, $\gamma$, and $\delta$ by computing the inverse of the Hessian of the loss function at the estimated minimum, and then taking the diagonal terms.  These terms approximate the variances in parameter fits when the loss function is approximately Gaussian near the minimum \citep{bevington1992}.

\subsubsection{Computing $v_*$, $r_*$, and $C_{bc}$}


When measurements are obtained at two wavelengths, $\lambda_1$ and $\lambda_2$, the ice volume fraction $v_*$ and black carbon mixing ratio $C_{bc}$ can be extracted from the decay parameters $\beta_1$ and $\beta_2$.  Each term $\beta_i$ can be expressed as a function of $v_*$ and $C_{bc}$ by taking the product of Eqs.~\ref{eqn:c} and \ref{eqn:muatot}.  This results in a set of two equations which can be solved, first, for $v_*$:

\begin{equation}
    \label{eqn:v}
    v_* = \frac{b_2\beta_1 - b_1\beta_2}{c_0\left(a_1b_2 - a_2b_1\right) - d_1b_2\beta_1 + d_2b_1\beta_2} .
\end{equation}

For notational simplicity we have made the substitutions $a_i = B\Gamma_i$, $b_i = \rho_{ice}\textrm{MAE}_{bc}(\lambda_i)$, and $d_i = n_{ice}(\lambda_i)B - 1$. As before, the term $c_0$ refers to the speed of light in air.

Once $v_*$ has been obtained, $C_{bc}$ can be computed as follows:

\begin{equation}
    \label{eqn:Cbc}
    C_{bc} = \frac{1}{c_0b_1(1+f v_*)}\left[ \left( \frac{1}{v_*} + d_1\right)\beta_1 - c_0a_1 \right] = \frac{1}{c_0b_2(1+f v_*)}\left[ \left( \frac{1}{v_*} + d_2\right)\beta_2 - c_0a_2 \right] .
\end{equation}


Here we have subsituted $f = B-1$.  After $v_*$ and $C_{bc}$ have been computed, the grain radius $r_*$ can be computed from the spatial spread parameter $\gamma_i$ at either wavelength:

\begin{equation}
    \label{eqn:r}
    r_* = e_i \left[ \frac{2c_0}{3\gamma_iv_* (1 + d_iv_*)} -a_i - b_iC_{bc}(1+f v_*) \right]^{-1} .
\end{equation}

Here $a_i$, $b_i$, and $d_i$ and $f$ are defined as they were previously, and $e_i = 3(1-g)/2$.  Because $r_*$ can be computed using either $\gamma_1$ or $\gamma_2$, we evaluate Eq.~\ref{eqn:r} at both wavelengths, and then take the uncertainty-weighted average of the two values obtained in this way to arrive at our final estimate for $r_*$.  Uncertainties in $v_*$, $r_*$, and $C_{bc}$ are obtained via error propagation from uncertainties in $\beta_1$, $\beta_2$, $\gamma_1$, and $\gamma_2$.

If it is known that absorption by light absorbing particles is small compared to absorption by ice grains, then $v_*$ and $r_*$ can be computed from the fit parameters extracted from single-wavelength measurements.  First $v_*$ can be computed from the exponential decay rate $\beta$, as follows:

\begin{equation}
    \label{eqn:v1}
    v_* = \frac{\beta}{ac_0 - \beta d} , 
\end{equation}

and then $r_*$ can be obtained from the spatial spread rate $\gamma$, and our estimate of $v_*$:

\begin{equation}
    \label{eqn:r1}
    r_* = e\left[ \frac{2c_0}{3\gamma v_* (1 + dv_*)} -a \right]^{-1} .
\end{equation}

Here $a$, $b$, $d$, and $e$ retain their meanings from Eqs.~\ref{eqn:v} and \ref{eqn:r}.

\subsubsection{Evaluation Using Simulated Measurements}

\begin{figure}
\centering{\includegraphics[width=0.9\linewidth]{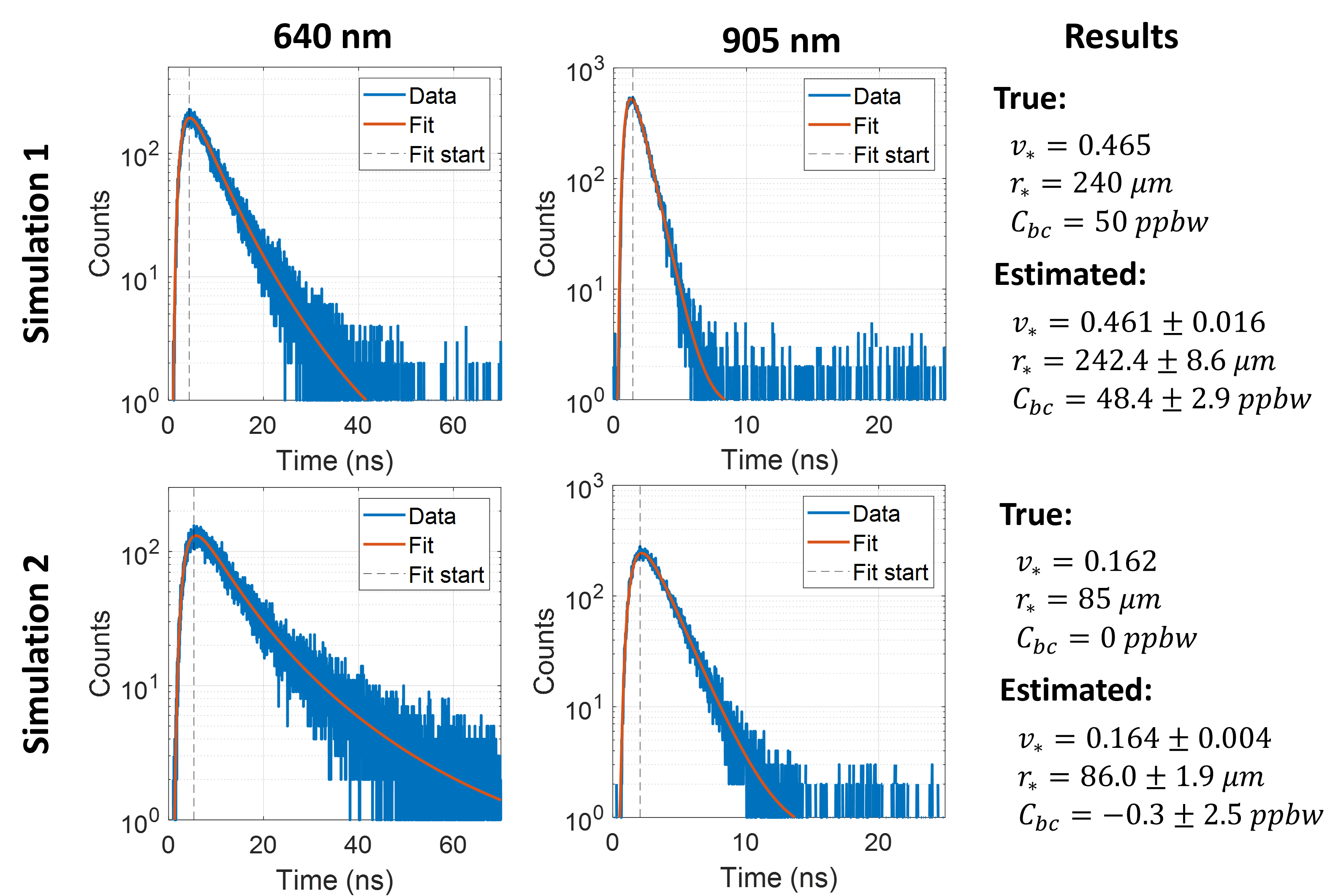}}
\caption{We used a monte-carlo photon transport simulator to validate our retrieval algorithm.  Measurements were simulated for 640 nm (left) and 905 nm (center) light for two simulated snow samples.  True and estimated snowpack parameters for each sample are shown on the right.}
\label{fig:mc}
\end{figure}

We validated our algorithm using a GPU-accelerated Monte-carlo photon transport simulation \citep{SnowLiDARMonteCarlo}, which was adapted from a simulator originally developed for tissue imaging studies \citep{satat2019}.  We modeled the propagation of photons within a semi-infinite, homogeneous scattering medium.  The medium's properties $\mu_a$, $\mu_s'$, and $c_*$ were computed from $v_*$, $r_*$, and $C_{bc}$ using Eqs.~\ref{eqn:muatot}, \ref{eqn:museff}, and \ref{eqn:c}.  To simulate pencil-beam illumination, photons were launched at the origin ($[x, y, z] = \mathbf{0}$) at time $t=0$ and at normal incidence to the snow surface.  Photons scattered randomly in the medium until they were absorbed, exited the medium, or satisfied an outlier termination criterion such as maximum number of scattering events.  For more details, we refer the reader to Chapter 4 of \cite{welch95}.

We simulated photon time-of-flight histograms at 640 nm and 905 nm measurement wavelengths for two snow samples with different properties.  We binned photons by the transverse position $s$ (bin width 1 cm) and time $t$ (bin width 16 ps) that photons exited the snow surface.  In the first simulation, for 640 nm measurements, photons detected at $s=8.0 \pm 0.5$ cm were used for curve fitting, whereas for 905 nm measurements photons detected at $s=5.0 \pm 0.5$ cm were used. In the second simulation, for 640 nm measurements, photons detected at $s=10.0 \pm 0.5$ cm were used for curve fitting, whereas for 905 nm measurements photons detected at $s=7.0 \pm 0.5$ cm were used.  Once a histogram of signal photons was created, a random number of background counts was added to each timing bin by sampling from a Poisson distribution with rate parameter $\eta$ that was chosen to be consistent with the uniform background count levels observed in experimental measurements.

Our results are shown in Fig.~\ref{fig:mc}.  For the first simulation, the true snowpack properties were $v_*$ = 0.465, $r_*$ = 240 $\upmu$m, and $C_{bc}$ = 50 ppbw.   Our method retrieved values of $v_*$ = 0.461$\pm$0.016, $r_*$ = 242.4$\pm$8.6 $\upmu$m, and $C_{bc}$ = 48.4$\pm$2.9 ppbw.  For the second simulation, the true snowpack properties were $v_*$ = 0.162, $r_*$ = 85 $\upmu$m, and $C_{bc}$ = 0 ppbw.   Our method retrieved values of $v_*$ = 0.164$\pm$0.004, $r_*$ = 86.0$\pm$1.9 $\upmu$m, and $C_{bc}$ = -0.3$\pm$2.5 ppbw.  These results suggest that our algorithm should produce accurate estimates under the idealized conditions prescribed here, and if our snow scattering model is correct.

\section{Materials}

\subsection{Apparatus}

\subsubsection{Lidar System}

\begin{figure}
\centering{\includegraphics[width=\linewidth]{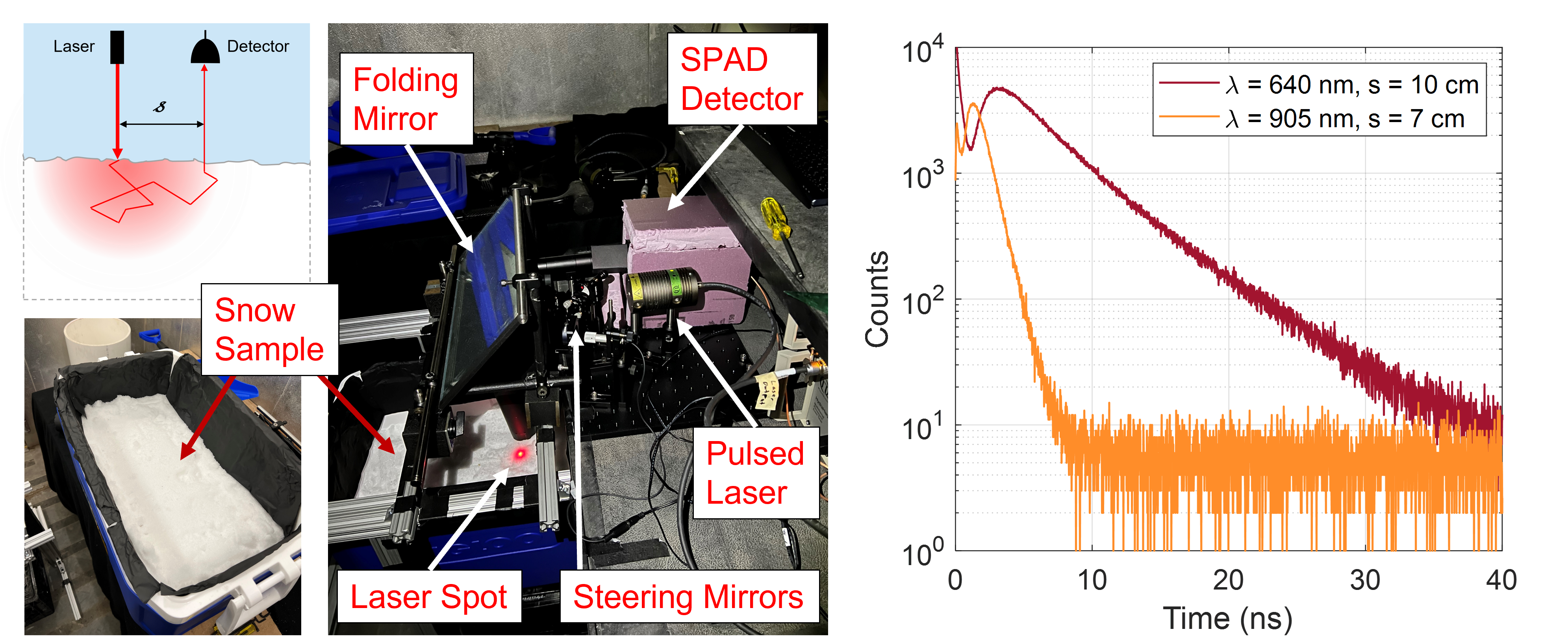}}
\caption{(Left) Photos depicting our experimental setup.  Snow was held in a cooler placed on the floor, and illuminated using pulsed diode lasers at two wavelengths (640 nm, 905 nm).  A single-photon avalanche diode (enclosed in pink insulating foam) measured the time-of-flight of photons that exited the snow surface at distance $s$ from the laser spot. (Right)  Time-of-flight histograms measured using our system.  For this test the snow sample consisted of natural snow that had been aged for nine months at $-10$ $^\circ$C.}
\label{fig:xphisto}
\end{figure}

We assembled a simple lidar system to measure the time-domain optical response of a variety of snow samples.  We perform time-correlated single-photon counting, in which a histogram of photon times of flight is built up over time by repeatedly illuminating the snow surface with a pulsed laser.  Photographs of our experimental setup are shown on the left in Fig~\ref{fig:xphisto}.  Our lidar system used a single-pixel SPAD detector (Microphoton Devices PDM series) with a timing jitter of $\sim$50 ps (FWHM), and two pulsed diode laser sources---a red laser with a wavelength of 640 nm (Picoquant LDH-P-C-640B), and a near-infrared laser with a wavelength of 905 nm (Picoquant LDH-P-C-905).  Each laser was operated at a pulse repetition frequency of 2.5 MHz and had a quoted pulsewidth of <90 ps.  The 640 nm laser was operated at a time-averaged power of 80 $\upmu$W, and the 905 nm laser was operated at a time-averaged power of 55 $\upmu$W.  A Picoquant Hydraharp 400 was used to synchronize the arrival times of detected photons with the laser repetition rate.  The overall instrument response function (IRF) of the system was measured to be 128 and 160 ps (FWHM) for 640 and 905 nm measurements, respectively.

\subsubsection{Measurement Procedure}

Experiments were conducted in a cold room at $-1$ $^\circ$C.  The room's lights were switched off and windows blacked out to reduce interference from ambient background light.  A large folding mirror was used to direct the lidar beam and detector field of view (FOV) towards a cooler filled with snow that was placed on the floor.  Because only a single laser diode could be operated at any one time, 640 nm measurements were collected first.  During these measurements, a red bandpass filter (Edmund Optics TECHSPEC 650nm/50nm) was placed in front of the detector to suppress interference from ambient background light.  Following these measurements the 640 nm laser head and bandpass filter were removed and replaced with the 905 nm laser head and a near-infrared bandpass filter (Thorlabs FL905-10).  We then collected a second set of measurements.

The beam from either laser head could be scanned by hand using a set of steering mirrors.  A lens was placed in front of the detector to focus its FOV to a small spot ($<1$ cm FWHM) on the snow surface.  To find this focus spot, the laser beam would be steered to the point on the surface at which detector counts were maximized.  Once the focus spot was found, a laser pointer (distinct from the pulsed diode lasers) was steered to mark the position of the focus spot.  The \emph{pulsed} beam could then be steered to a point on the snow surface that was displaced from the focus spot by a small distance $s$ that was measured using a ruler.  The focus-marker beam would then be switched off.  When the 905 nm laser was in use, a phosphorescent laser viewing card was used to find the position of the laser spot on the snow surface.

We note that even when the laser and focus spots were separated by several centimeters, interference from direct returns off of the snow surface remained significant due to phenomena such as lens flare.  Although we could not suppress this interference entirely, we were able to mitigate it by placing a long lens tube in front of our detector that functioned as a baffle.

For each snow sample, and for each laser wavelength, we collected measurements at multiple source-detector separations $s$.  Each measurement consisted of a histogram of photon arrival times with 16 ps timing bins that spanned a 250 ns timing window.  Examples of histograms measured by our lidar system are shown on the right in Fig.~\ref{fig:xphisto}.  The first measurement would always be collected at $s$ = 0 cm to measure the time-of-arrival of photons that scattered directly off of the snow surface.  The peak of this direct return would serve as a reference time for all subsequent measurements.  Direct surface returns were always measured with a 60 second integration time, with a neutral density filter placed in front of the detector to prevent saturation, and with a wooden ruler placed on the snow surface at the position of the laser spot to prevent bias due to subsurface scattering.  Following this, histograms would be collected for one or more non-zero source-detector separations.  We used an integration time of 10 minutes for each histogram collected with 640 nm light, and 30 minutes for each histogram collected with 905 nm light.  A longer integration time was required at 905 nm because our SPAD detector was less sensitive at this wavelength, the output power of our laser was lower, and the snow itself was less reflective.  Ice grains from each sample were inspected before and after each set of measurements to ensure that snow properties had not changed significantly due to metamorphism. 

Before proceeding, we want to stress that our lidar system was assembled strictly for the proof-of-principle demonstrations documented in this paper.  It was not optimized for ease of use or light collection efficiency.  Although the integration times reported here are quite long, we expect that a cleverly engineered system might collect equivalent data with integration times that are far shorter---perhaps by several orders of magnitude.  Integration time could be reduced significantly, for instance, by using a multipixel Silicon Photomultiplier (SiPM) in place of the single-pixel SPAD used here, and by using lasers with higher power and higher repetition rates.  The use of laser sources and SPADs designed for a consumer electronics environment \citep{king2023}, rather than the optical bench equipment used here, would also allow for a system that was portable, rugged, and affordable.  Altogether, this suggests that the development of a field-deployable system is a feasible goal---one which we hope to pursue in future work.

\subsection{Samples}

We performed two sets of experiments.  In the first, samples had relatively low LAP concentrations but grain size and density varied significantly.  In the second, the samples had varying amounts of black carbon mixed into them, but density and grain size was relatively constant.

All snow used in our experiment originated as natural snow harvested on Dartmouth College campus and was subsequently modified in various ways.  When not being used for experiments, snow samples were stored in lidded coolers in a $-10$ $^\circ$C cold room.

\subsubsection{Clean Snow Samples}
\label{sec:cleansample}

We performed five sets of measurements on samples with varying density and grain size but relatively low LAP content.  The snow used in the first set of measurements was harvested after a snowfall in March 2022 and then kept in a $-10$ $^\circ$C cold room for nine months.  By the time measurements were taken, the snow had become more dense and the grains had metamorphosed into medium size rounded grains and rounding faceted particles \citep{fierz2009}.  The next three data collections were performed on a single snow sample that was modified between measurements.  The sample was harvested 30 minutes after snow had ceased falling, immediately outside our laboratory at Dartmouth College.  It was then stored overnight at $-10$ $^\circ$C.  Measurements were collected the next morning on the unmodified sample, which had a very low density and consisted of precipitation particles \citep{fierz2009}, with many stellar dendrites.  A second set of measurements was collected after the snow had been compacted with a shovel---thus increasing its density but leaving grain size and shape relatively unchanged.  The third set of measurements was collected after the snow was aged for three weeks at $-10$ $^\circ$C and then for one day at $0$ $^\circ$C.  This aging produced a clear change in grain shape, to small rounded grains and decomposing precipitation particles, and a small increase in grain size and density.  For our final set of measurements we harvested snow that had been sitting outside for weeks, where it had experienced several melt and re-freeze events.  This snow had very high density and coarse grains.

At the time of data collection, all samples were held in coolers with approximate internal dimensions of 50 cm$\times$25 cm$\times$30 cm and that had matte white internal walls.  Snow would fill the cooler to varying degrees, but was typically at least 20 cm deep, relative to the cooler bottom.

\subsubsection{Soot Addition Experiments}
\label{sec:sootsample}

For the second set of experiments, we filled five Styrofoam coolers (dimensions 17.5 cm$\times$23.5 cm$\times$24.0 cm) with freshly fallen snow.  We then mixed small amounts of Sigma-Aldrich Fullerene Soot (572497) into the samples, such that the five respective samples had 0, 1, 2, 3, and 4 baseline units of soot.  To add soot to the snow in a controlled fashion, we created a soot-water suspension with a known concentration of soot, and then applied controlled volumes of the suspension to each snow sample with a spray bottle.  The soot was mixed evenly into the snow using an ice scraper.

After performing a first set of measurements on the sooty samples we found that the added soot had a weaker effect on the snowpack absorption coefficients than had been expected.  Following this finding, we approximately doubled the added soot concentration in all samples and repeated the measurements.


\subsection{Ground Truth Measurements}

Ground truth ice volume fraction was measured by extracting a small core (depth $\sim$5 cm) from the snow surface.  We measured the volume of snow in the core.  The snow was then allowed to melt, and we measured the volume of the meltwater.  Ice volume fraction was computed from the snow and meltwater volumes using conservation of mass.

Ground truth grain size was measured by imaging a small, snow-filled test tube (1.4 cm internal diameter) with a SkyScan 1172 microCT scanner (40 kV, 250 $\upmu$A source, 17 $\upmu$m resolution).  Bruker NRecon software was used to reconstruct a 3D image of the sample.  Following guidance from \cite{hagenmuller2016}, the image was then blurred with a Gaussian kernel (radius 1 pixel), binarized with Otsu's method, and morphologically ``opened'' (radius 1 pixel).  The surface area to volume ratio (SA/V) of the imaged sample was then computed two times using Bruker's CTAN software, following marching squares (2D analysis) and marching cubes (3D analysis) surface reconstructions.  We computed the grain radius from each SA/V ratio independently, and then used the average of these two values as ground truth.


Ground truth estimates of black carbon concentration were obtained using a single particle soot photometer (SP2; Droplet Measurement Technologies), in a manner similar to that reported in \cite{lazarcik2016}. Each snow sample was melted and ultrasonicated for at least 15 minutes prior to analysis. The liquid snow samples were aerosolized using an ultrasonic nebulizer (CETAC U5000AT), which removes moisture from the liquid stream before passing aerosols such as black carbon onto the SP2. The SP2 estimates black carbon particle mass via measurements of laser-induced incandescence. This system was calibrated using a series of fullerene soot standards. To avoid saturating the SP2, snow samples that were expected to have particularly high black carbon concentrations were diluted with MilliQ water by a factor of 6.

\section{Results}

\subsection{Clean Snow Experiments}

\subsubsection{Individual Snow Sample}

To provide insight into our data collction and fitting procedures, we first present a detailed review of all measurements collected for a single snow sample.  This sample, which is described in greater detail in the Materials section, consisted of natural snow that had been aged for nine months in a $-10$ $^\circ$C cold room.  

\begin{figure}
\centering{\includegraphics[width=\linewidth]{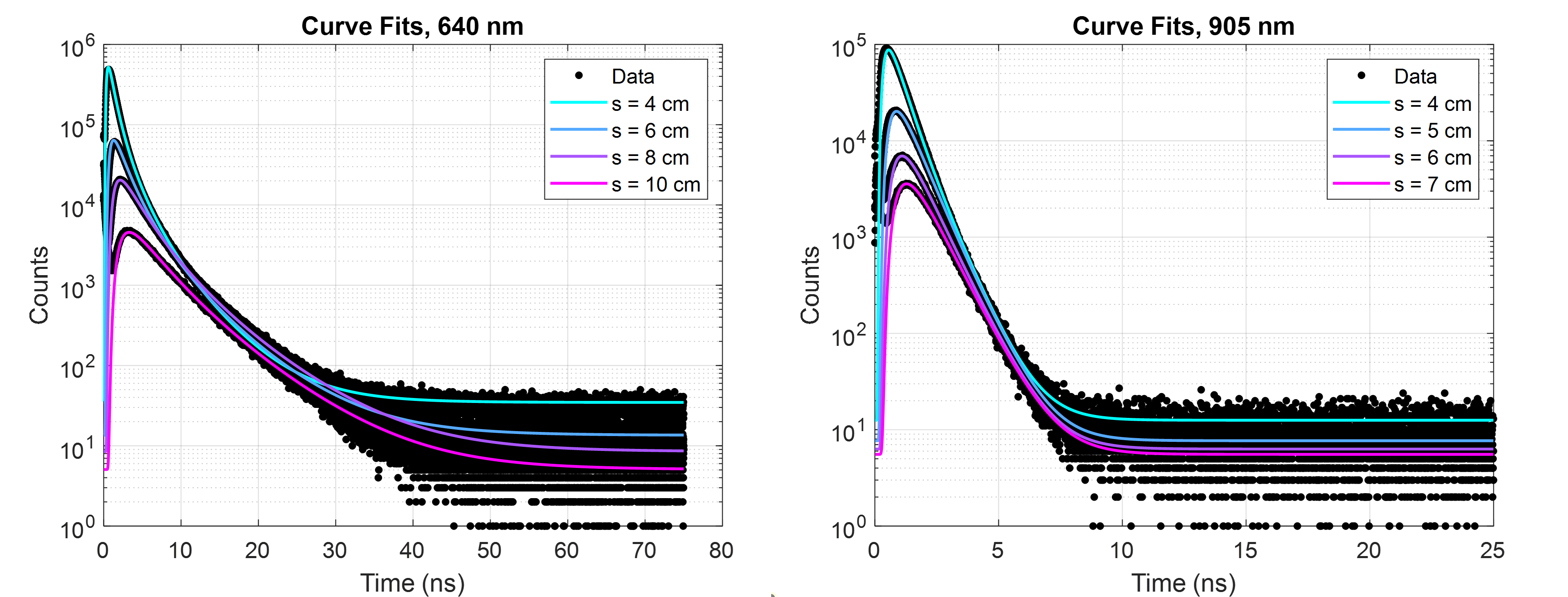}}
\caption{Raw measurements collected for a single snow sample.  Time-of-flight histograms were measured at two wavelengths (640 nm, 905 nm) and for four source-detector separations per wavelength.  A diffusion model was fit to each histogram.  The pair of curves with the best goodness of fit was used to compute snowpack properties.}
\label{fig:data0117}
\end{figure}

The raw, time-of-flight histogram data collected for this sample, as well as our curve fits to those measurements, are shown in Fig.~\ref{fig:data0117}.  Measurements were taken at four different source-detector separations for each wavelength: $s$ = 4, 6, 8 and 10 cm at 640 nm, and $s$ = 4, 5, 6 and 7 cm at 905 nm.  As a rule of thumb, we would start each fit at a timing bin that corresponded to the peak of the diffusion signal.  This was done to avoid fitting to the earliest arriving photons, which are poorly described by our diffusion model.  Histograms were collected at multiple $s$ values because it was not known \emph{a priori} what range of $s$ values would yield good diffusion curve fits.  If $s$ and $\mu_s'$ were both small, then photons in the signal peak would be poorly described by our diffusion model because they would exit the snowpack after too few scattering events.  On the other hand, if $s$ and $\mu_s'$ or $\mu_a$ were too large, the diffusion signal would be faint relative to background interference, and the fit would be poor.

\begin{figure}
\centering{\includegraphics[width=\linewidth]{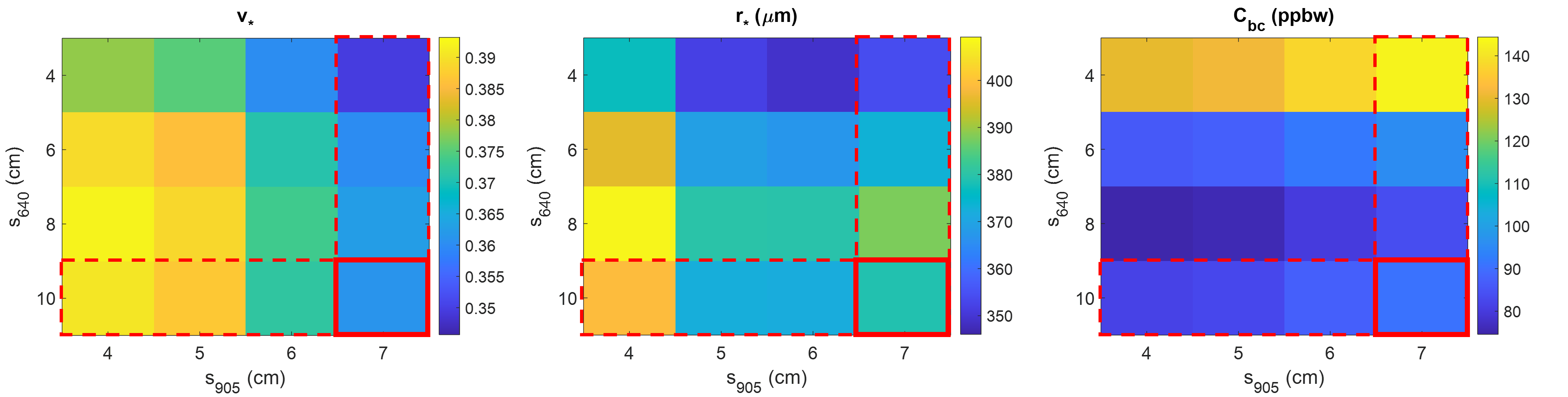}}
\caption{Dependence of $v_*$, $r_*$, and $C_{bc}$ estimates on choice of source-detector separation $s$ for each measurement wavelength.  Estimates that correspond to the pair of curve fits with the lowest reduced deviance \citep{mccullagh2019} are outlined in red.}
\label{fig:vary0117}
\end{figure}

In Fig.~\ref{fig:vary0117}, we show how the retrieved snow properties varied with respect to our choices of source-detector separation at each wavelength.  In general, estimates of $v_*$, $r_*$, and $C_{bc}$ did not vary very much so long as good curve fits were obtained at both wavelengths, but diverged from the typical value when one or both of the curve fits were poor.  As an example, it is evident in Fig.~\ref{fig:vary0117}(c) that $C_{bc}$ estimates are biased high at $s_{640}$ = 4 cm, but are otherwise relatively insensitive to changes in $s$ at either wavelength.

To arrive at a single estimate for $v_*$, $r_*$, and $C_{bc}$, we chose the curve fit at each wavelength with the lowest \emph{reduced deviance} \citep{mccullagh2019}.  Deviance is a goodness of fit metric that is appropriate for data that follows Poisson statistics, and that is asymptotically equivalent to $\chi^2$ goodness of fit when the number of counts in all histogram bins is high.  For the data collection described here, the best fits corresponded to $s$ = 10 cm at 640 nm and $s$=7 cm at 905 nm.  From the parameters of these two fits we estimated that $v_*$ = 0.361$\pm$0.004, $r_*$ = 379.0$\pm$4.1 $\upmu$m, and $C_{bc}$ = 90.7$\pm$1.3 ppbw.  As described previously, the reported uncertainties correspond to statistical uncertainties in the curve fit parameters, propagated through Eqs. \ref{eqn:v}, \ref{eqn:Cbc}, and \ref{eqn:r}. They do not account for potential inaccuracies in the diffusion or scattering models.  The ground truth measurements of $v_*$, $r_*$, and $C_{bc}$ were 0.465, 242.5 $\upmu$m, and 30.7 ppbw, respectively.

\subsubsection{Full Results}

\begin{figure}
\centering{\includegraphics[width=0.5\linewidth]{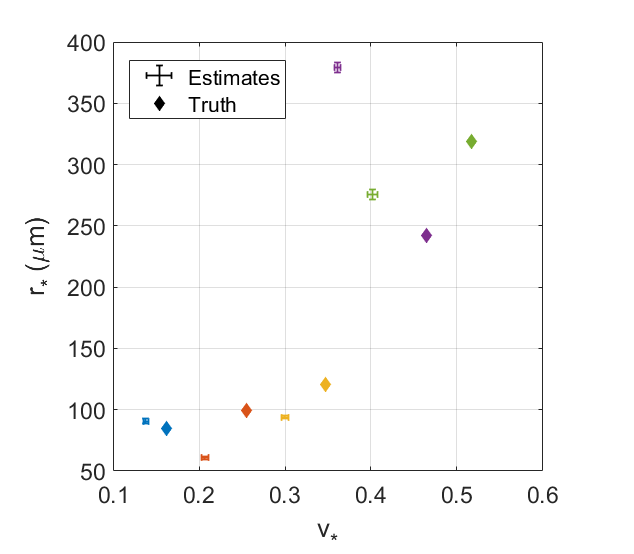}}
\caption{Summary of the estimated and ground truth ice volume fractions $v_*$ and grain sizes $r_*$ of five clean snow samples.  Error bars indicate one standard deviation.  Estimates and ground truth values are matched by color.}
\label{fig:vrscat}
\end{figure}

We now present a summary of all results obtained for the clean snow samples.  The properties of the snow samples used in these tests varied widely, from light, fine-grained fresh powder to dense, coarse-grained snow that had experienced several melt and re-freeze events.  In Fig.~\ref{fig:vrscat}, we show a scatter plot of the densities and grain sizes estimated using our method, as well as ground truth values.  

\begin{figure}
\centering{\includegraphics[width=\linewidth]{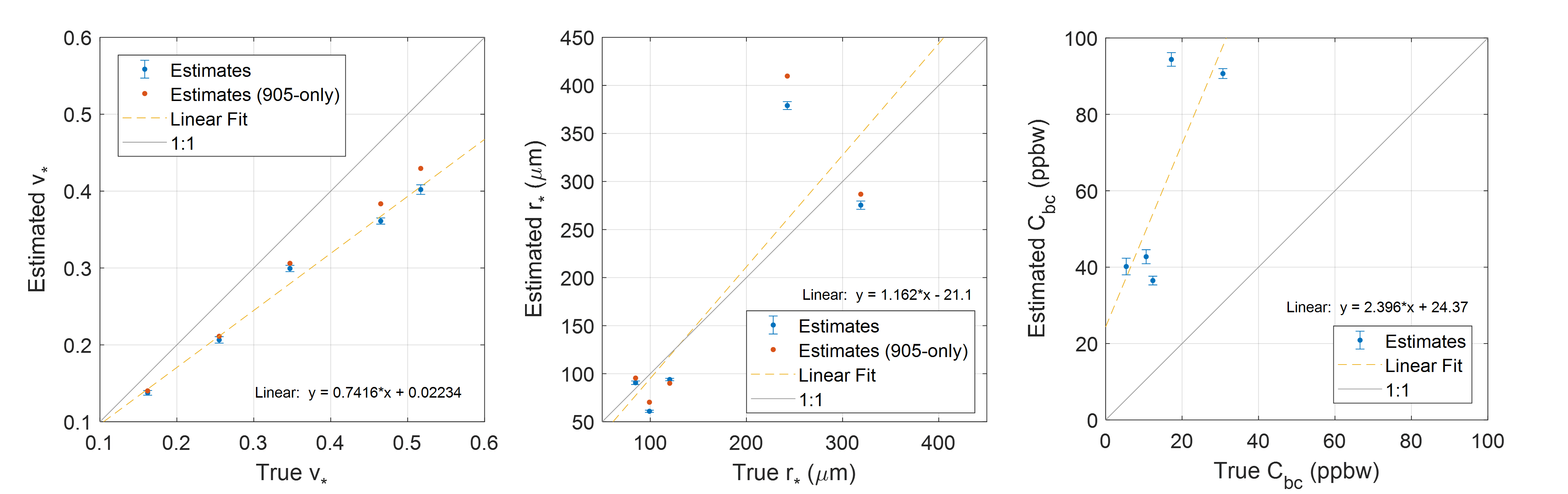}}
\caption{Ground truth versus estimated values of ice volume fraction $v_*$, grain size $r_*$, and black carbon mass mixing ratio $C_{bc}$ for five clean snow samples.  Blue marks indicate estimates obtained using two measurement wavelengths (640 nm, 905 nm).  Error bars indicate one standard deviation.  Red marks indicate estimates of $v_*$ and $r_*$ obtained from 905 nm measurements only.  Uncertainties were not computed for 905-only estimates.}
\label{fig:cleansnowprops}
\end{figure}

Our estimates of $v_*$, $r_*$, and $C_{bc}$ are plotted with respect to ground truth in Fig.~\ref{fig:cleansnowprops}.  In Fig.~\ref{fig:cleansnowprops}(a), we see a clear positive and nearly linear relationship between the ice volume fraction estimated using our technique, and ground truth, although estimates appear to be biased towards lower densities.  The trends for $r_*$ and $C_{bc}$ are less clear, although our method appears to be capable of distinguishing between small and large grain sizes, and low and moderate impurity concentrations.  To the extent that trends can be observed, there appears to be an approximately 1:1 relationship between $r_*$ estimates and ground truth, whereas $C_{bc}$ appears to be over-estimated by a factor of $\sim$2.5.  All statistical uncertainties in $v_*$ and $r_*$ estimates are 1-2\% of the estimated value.  All statistical uncertainties in $C_{bc}$ estimates are 1-2 ppbw.  Notably, these uncertainties are comparable to the statistical uncertainties reported by \cite{allgaier2022ice} in their estimates of black carbon concentrations in glacier ice.

In addition to dual-wavelength estimates of $v_*$, $r_*$, and $C_{bc}$, we also show estimates of $v_*$ and $r_*$ computed using only 905 nm measurements.  The single-wavelength results match the dual-wavelength results very closely.  Ice volume fraction estimates are slightly higher, which is consistent with excess absorption due to unmodeled LAPs.  Single-wavelength grain size estimates are alternately higher or lower than corresponding dual-wavelength estimates.

Considering the very small statistical uncertainties in our results, we expect that the biases seen here are most likely attributable to model mismatch.  In particular, the excess black carbon content predicted by our method is plausibly explained by the presence of other kinds of light absorbing impurities such as dust.  The samples used in this test were collected outdoors and were handled with shovels, ice scrapers, and various other equipment that may have been coated with dust or dirt.  Further investigation is needed to understand the relatively small bias in estimates of $v_*$.  One possible explanation is that the measured signal was influenced by unmodeled reflections from the white side-walls of the cooler.  A deeper analysis would be required to confirm this.  However, one would expect such reflections to reduce the observed decay rate by scattering photons back into the probed region instead of allowing them to escape.  The effect of ``close packing'' of grains has been reported to reduce the absorption enhancement factor of snow \citep{liboisB}.  However, it is not clear that the snow samples used in this study were close-packed to an extraordinary degree.  There is a notable outlier among the $r_*$ estimates that is approximately 50\% higher than its ground truth measurement.  The origins of this outlier are unclear.   By inspection of Eq.~\ref{eqn:museff}, we see that the estimated $r_*$ value would be reduced by 50\% if the modeled scattering asymmetry was increased from 0.825 to 0.883. It is thus possible that the outlier snow sample---which consisted of medium size rounded grains and rounding faceted particles that had been aged for nine months in a $-10$ $^\circ$C cold room---had a higher scattering asymmetry than the others. However it is not clear why this would be so.



\subsection{Soot Addition Experiments}

Here we present the results of the soot addition experiments described in the Materials section, where the snow samples contained varying concentrations of black carbon.  For these tests, the source-detector separation was held fixed at $s$ = 8 cm for 640 nm measurements. 
 For 905 nm measurements, a value of $s$ = 5 cm was typically used, although this was occasionally reduced to 4 cm if the measured signal would otherwise be too faint to yield a good curve fit.

\begin{figure}
\centering{\includegraphics[width=0.6\linewidth]{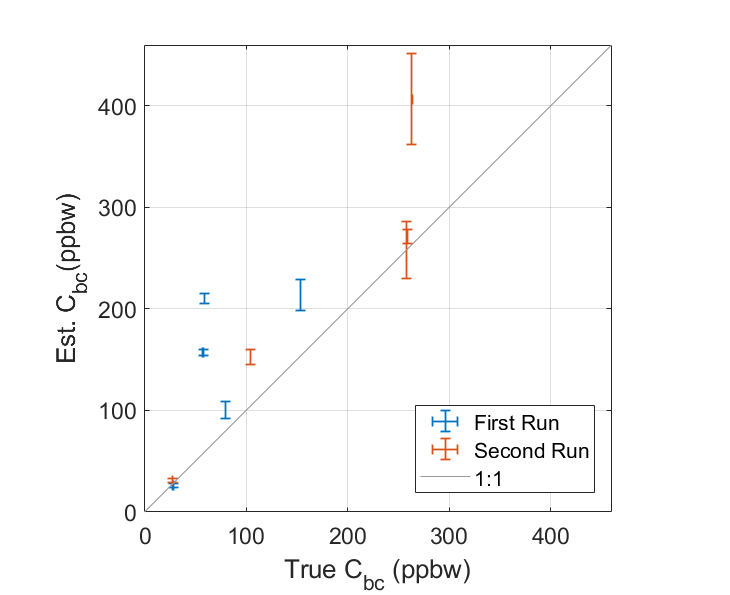}}
\caption{Retrieved values of black carbon mass mixing ratio $C_{bc}$ plotted with respect to ground truth estimates obtained by a single-particle soot photometer (SP2).  Increasing quantities of Sigma-Aldrich fullerene soot were added to five snow samples.  Results from the first set of measurements are shown in blue.  Soot concentrations were then approximately doubled for all samples and a second set of measurements was taken.  Results from the second set of measurements are shown in red.  Error bars indicate one standard deviation.  SP2 errors are typically too small to be visible.}
\label{fig:Cbcest}
\end{figure}

 The primary goal of these experiments was to assess the accuracy and sensitivity of the estimates of black carbon mass mixing ratio produced by our method.  To this end, in Fig.~\ref{fig:Cbcest} we show a plot of the $C_{bc}$ values estimated with our method versus ground truth estimates obtained using an SP2.  Blue data points correspond to the first set of measurements, for which the soot concentrations were relatively low, and red data points correspond to a second set of measurements that was collected after the added black carbon concentration in each snow sample  had been approximately doubled.

 Upon inspection we see a clear correlation between the estimated and ground truth $C_{bc}$ values.  The correlation is approximately linear and nearly one-to-one.  Two outlier data points (with ground truth $C_{bc}$ of 58, 59 ppbw) lie off of the one-to-one line.  We expect that the outliers are the result of an error in the ground truth estimates.  It is possible that our mixing process did not uniformly distribute the black carbon content throughout the snow and that the region sampled for SP2 analysis was unusually clean.

\begin{figure}
\centering{\includegraphics[width=\linewidth]{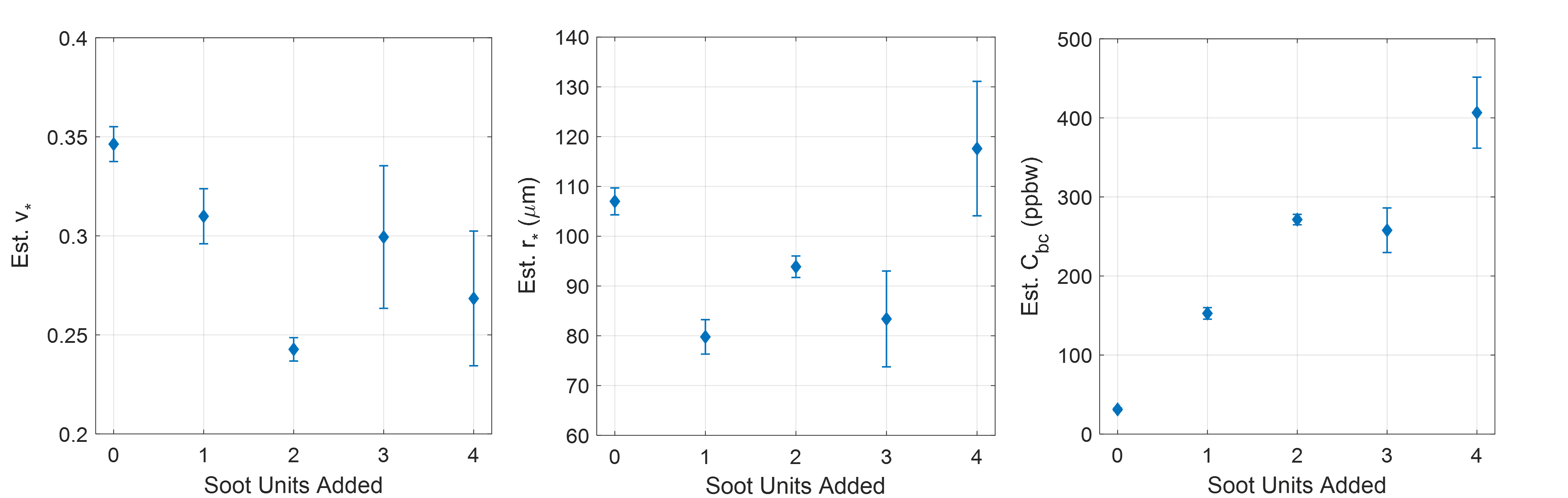}}
\caption{Estimates of $v_*$, $r_*$, and $C_{bc}$ obtained for ``double soot'' samples.  Estimates are plotted as a function of the number of soot units mixed into the snow.  One soot unit corresponds to a fixed volume of soot-water suspension that is applied to the snow sample with a spray bottle and then mixed into the snow volume.  Error bars indicate one standard deviation.}
\label{fig:vrcsooty}
\end{figure}

For good measure, we also show the estimates of ice volume fraction, grain radius, and black carbon concentration obtained for the five ``double-soot'' samples described previously.  Estimates are plotted in Fig.~\ref{fig:vrcsooty} as a function of total units of soot-water solution that were applied to each sample using a spray bottle.  Although ground truth $v_*$ and $r_*$ were not collected, results in Fig.~\ref{fig:vrcsooty}(a) and (b) indicate that the density and grain size of the snow samples was relatively consistent, but did have some variance.  This variance may have been caused by differences in how each sample was mixed, or from interaction with the liquid water in the soot-water suspension.  Regardless, we see in Fig.~\ref{fig:vrcsooty}(c) that estimated $C_{bc}$ increases approximately linearly and nearly monotonically as a function of the amount of added soot, with no clear dependence on density or grain size.

\section{Discussion}

In this work we have introduced a new method for measuring the density, grain size, and black carbon content of a dry snowpack using non-invasive, time-domain diffuse optical measurements.  We have presented a model for the time-domain optical response of a snowpack that was adapted from the biomedical optics literature \citep{kienle1997, haskell1994}.  Our model was obtained by solving the photon diffusion equation---an approximation to the radiative transfer equation that accurately describes the propagation of light in highly scattering media \citep{welch95}.  We used a geometric scattering model to relate the parameters of our photon diffusion model to dry snowpack density, grain size, and black carbon concentration.  Our scattering model was derived from a well-known snow-optics model \citep{kokh2004}, but extended to account for the effective speed of light within snow, as well as the mixing state of black carbon.  We then developed an algorithm to retrieve the snowpack properties from time-domain optical measurements collected at two wavelengths.

We were able to validate our method in a series of proof-of-principle experiments in which we measured the properties of real snow samples using a photon-counting lidar system.  The results of these experiments are encouraging.  We see a clear, nearly linear correlation between the snowpack densities estimated by our method, and ground truth.  When the LAPs in the snow were known to be black carbon particles, we observed a nearly one-to-one correlation between the black carbon mass mixing ratios estimated using our method, and those measured using a single-particle soot photometer \citep{schwarz2006}.  A nearly one-to-one correlation was also found between the grain sizes measured by our method and those determined from micro-CT images---although this correlation was not as strong.  Our goal in this work was to obtain proof-of-principle results.  More experiments are required to comprehensively assess our method's accuracy, biases, and failure modes.

Although our results are encouraging, we believe that the primary contribution of our work is not necessarily the exact method that we have proposed, but rather that we have been able to clearly demonstrate that time-domain diffuse optics is an appropriate sensing modality for measuring snowpack properties. 
Previous works have used ray-tracing simulations to explore the feasibility of inferring snow properties from time-domain diffuse optical signals \citep{libois2019}, and to predict the relationship between snow properties and lidar altimetry biases \citep{smith2018}.  However, as far as we are aware, our work is the first to provide clear \emph{experimental} evidence that the optical response of a snowpack that has been illuminated by a laser pulse can be accurately described using a photon diffusion model; and also that this response is measurably influenced by changes to important snowpack properties like grain size, density, and impurity content.

Our method could be improved in several ways.  More sophisticated models that incorporate features such as liquid water content in the snow, finite snow depth, or surface roughness might be developed to enable retrieval of snow properties in a broader set of circumstances.  Our measurement procedure could also be improved.  In our experiments, processing a single snow sample required between 40 minutes to several hours of data collection time.  This could be dramatically reduced by improving our instrument design to incorporate multi-pixel SPAD detectors, higher power lasers, or by simply placing the laser and detector closer to the snow surface.  The integration times used in our experiments were also conservative---further analysis could determine the minimum number of photons required to accurately retrieve snow properties.  Finally, using our method in the field would require the development of a rugged and portable instrument.  The dramatic decrease in the cost and size of pulsed lasers and photon counting detectors in recent years makes this possible.  All components required for such a device can be found in the current model of the iPhone Pro \citep{king2023}.

Our work provides what is, to our knowledge, the first experimental demonstration of snow density estimation from non-invasive optical measurements.  Previous works used ray-tracing simulations to demonstrate non-invasive porosity ($=1-v_*$) measurements in arbitrary porous media \citep{libois2019}, or estimated snow density using invasive optical transmission measurements \citep{gergely2010}.  Our method could potentially provide field measurements of snow-water-equivalent (SWE)---the product of snow depth and density---or surface density, which might in turn prove useful in hydrological or ecological studies, or for validating remote sensing techniques \citep{kinar2015}.  Our method might also enable field measurements of the concentrations of light absorbing impurities in snow.  Current methods for quantifying LAP concentrations require collecting samples and transporting them offsite for spectrophotometric \citep{grenfell2011}, thermo-optical \citep{lazarcik2016}, or chemical \citep{lawrence2010} analysis.  Our method could be made more sensitive to LAP absorption through the use of shorter wavelength (e.g.~blue or green) illumination, for which ice is less absorptive \citep{warren2008}.  Using more than two wavelengths could permit simultaneous retrieval of the concentrations of different LAP types---such as black carbon and dusts with different mineral compositions \citep{moosmuller2009}. 

Although the instrument used in this study was assembled from the same components that make up a typical photon counting lidar system, our measurements were effectively \emph{in situ} because our lidar was always placed within a meter of the snow's surface.  In the future, we hope to  develop true \emph{remote} lidar sensing techniques that are grounded in time-domain diffuse optics models.  The leap from in situ to remote measurements poses new challenges that include dramatically lower photon counts, wider beam footprints, and confocal measurement geometries.  Further analysis is required to determine which snow properties can be feasibly retrieved under these conditions.  The ability to remotely measure snow density could in turn enable remote mapping of SWE.  Radar retrievals of SWE have been proposed \citep{tsang2022, deeb2012}, but can be thwarted by tree cover or by the presence of liquid water in snow \citep{deeb2012}.  In comparison, the ability to penetrate tree canopies is a well known feature of lidar scanners \citep{deems2013}, and the optical similarity of liquid water and ice \citep{warren2019} suggests that lidar retrievals should be robust to small concentrations of liquid water in snow.  On the other hand, compared to radar, the smaller penetration depth of diffuse optical signals may limit lidar SWE retrievals to shallow snowpacks.  The ability to remotely map LAP concentrations using lidar would, among other things, enable more accurate predictions of snow radiative forcing \citep{painter2013}.  Algorithms that retrieve LAP concentrations from hyperspectral imagery have been proposed \citep{zege2011}, however it has been shown that such retrievals are only practical for exceptionally dirty snow \citep{warren2013}.

An alternative direction for future work is the development of more advanced algorithms for processing in situ measurements that leverage decades of advances in diffuse optical spectroscopy research \citep{sekar2019}.  In particular, the adaption of \emph{diffuse optical tomography} methods \citep{okawa2023} to snow would enable non-invasive retrieval of snow stratigraphy, or even full 3D mapping of snowpack properties within a probed region.  Observations of snow stratigraphy are often made to assess the structural stability of the snowpack to predict avalanche risk, as well as the history of snow deposition and metamorphism \citep{nienow2011}.  

Finally, time-domain optical measurements represent a new opportunity to develop a deeper understanding of the scattering optics of snow.  Close examination of time-resolved measurements may yield insights that facilitate improvements to snowpack albedo models.  It's also possible that more sophisticated  measurements, such as the time-resolved measurement of light penetration or bidirectional reflectance, are sensitive to snowpack properties that have been challenging to measure by other means---such as scattering asymmetry or the absorption enhancement factor.

\section{Acknowledgements} 

We wish to thank our reviewers for their helpful comments and discussion.  We thank Brent Minchew and Joanna Millstein for the discussions that inspired this project and for support with initial theoretical work.  Andrii Murdza provided general assistance with micro-CT measurements and cold room experiments.  Anna Valentine prepared the snow samples used in our soot addition experiments.  Jacob Chalif ran our SP2 measurements and helped us interpret the results.  Connor Henley was supported by a Draper Scholarship and by a grant from the Office of Naval Research (N00014-21-C-1040). Any opinions, findings and conclusions or recommendations expressed in this material are those of the author(s) and do not necessarily reflect the views of the Office of Naval Research.  Colin R. Meyer was supported by the Heising-Simons Foundation (\#2020-1911), the Army Research Office (78811EG), and the National Science Foundation (2024132).


\bibliography{references}   
\bibliographystyle{igs}  





\end{document}